\documentclass[nosubsecnum]{svmult}
\usepackage{graphicx}  
\usepackage{subfigure}   
 \usepackage{epsfig}          
\usepackage{epstopdf}
\usepackage{mathptmx,helvet,courier}  
\usepackage{multicol}        
\usepackage[bottom]{footmisc} 
\usepackage{cite}
\usepackage{amsmath}\usepackage{amssymb}

\def\rcgindex#1{\index{#1}}
\def\myidxeffect#1{{\bf\large #1}}

\begin{document}


\title*{A supersonic crowdion in mica}
\subtitle{Ultradiscrete kinks with energy between $^{40}$K recoil
and transmission sputtering}
 \titlerunning{A supersonic crowdion in mica}
 \author{Juan F. R. Archilla
\and Yuriy A. Kosevich \and No\'e Jim\'enez \and  V\'{\i}ctor J.
S\'{a}nchez-Morcillo \and Luis M. Garc\'{i}a-Raffi}
\authorrunning{JFR Archilla, YuA Kosevich, N Jim\'enez, VJ S\'{a}nchez-Morcillo and LM Garc\'{i}a-Raffi}
\tocauthor{J.F.R. Archilla, Yu.A. Kosevich, N. Jim\'enez, V.J.
S\'{a}nchez-Morcillo and L.M. Garc\'{i}a-Raffi}
\institute{J.F.R.~Archilla
 \at Group of Nonlinear Physics,  Universidad de
Sevilla, ETSII,
  Avda Reina Mercedes s/n, 41012-Sevilla, Spain,
 \email{archilla@us.es}
 \and Yu. A. Kosevich
 \at Semenov Institute of Chemical Physics, Russian
Academy of Sciences.  Kosygin street 4, 119991 Moscow, Russia,
 \email{yukosevich@gmail.com}
 \and N. Jim\'enez \at Instituto de Investigaci\'{o}n para la Gesti\'{o}n
Integrada de las Zonas Costeras, Universidad Polit\'{e}cnica de
Valencia, C/.Paranimfo 1,  46730 Grao de Gandia, Spain,
\email{nojigon@epsg.upv.es}
 \and V. J. S\'{a}nchez-Morcillo
 \at Instituto de Investigaci\'{o}n para la Gesti\'{o}n
Integrada de las Zonas Costeras, Universidad Polit\'{e}cnica de
Valencia, C/.Paranimfo 1,  46730 Grao de Gandia, Spain,
\email{victorsm@upv.es}
 \and L.M. Garc\'{i}a-Raffi
 \at Instituto Universitario de Matem\'{a}tica Pura y
Aplicada, Universidad Polit\'{e}cnica de Valencia, Camino de Vera
s/n, 46022 Valencia, Spain, \email{lmgarcia@mat.upv.es}
 } 

 \maketitle

\vspace{-2cm} \setcounter{minitocdepth}{1} \dominitoc

 \abstract{In this work we analyze in detail the
behaviour and properties of the kinks found in an one dimensional
model for the close packed rows of potassium ions in mica
muscovite. The model includes realistic potentials obtained from
the physics of the problem, ion bombardment experiments and
molecular dynamics fitted to experiments. These kinks are
supersonic and have an unique velocity and energy. They are
ultradiscrete involving the translation of an interstitial ion,
which is the reason they are called {\em crowdions}. Their energy
is below the most probable source of energy, the decay of the
$^{40}$K isotope and above the energy needed to eject an atom from
the mineral, a phenomenon that has been observed experimentally
 }

\section{Introduction}
\begin{figure}[b]
\centerline{\includegraphics[width=9cm]{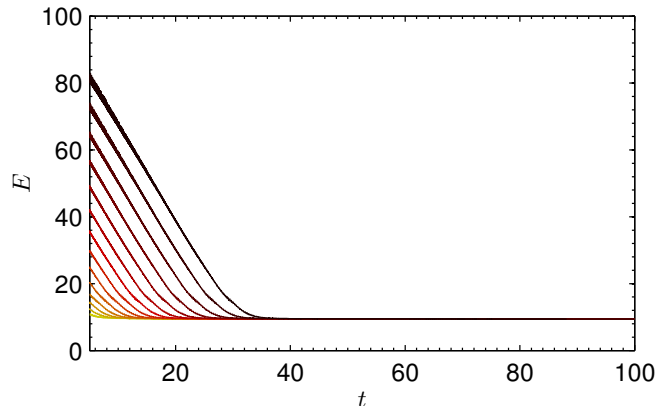}} 
    \caption{Energies of several kinks with respect to time. When more energy
    than the crowdion's one is delivered  and therefore a faster kink is produced,
    a radiation process takes place
    until the supersonic crowdion is formed. Thereafter, the crowdion is extremely
    stable. If the initial energy is smaller than the crowdion's one the kink dissipates into
    phonons. The scaled units are approximately 3\,eV for energy and 0.2\,ps for time.
    The final velocity and energy are approached asymptotically, being  $V_c=2.7387$ (7.2\,km/s) and
    $E_k=9.5$ (26.2\,eV) in scaled and physical units}
    \label{crowdion-figure01}
 \end{figure}

\begin{figure}[b]
\sidecaption[b]
\includegraphics[width=7cm]{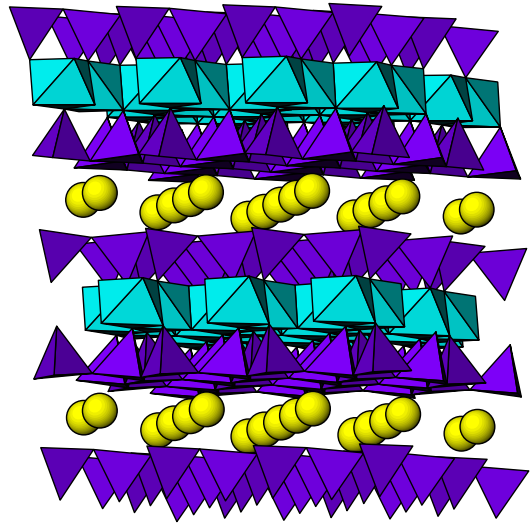} 
    \caption{The structure of mica muscovite where the potassium
    layer can be observed. This point of view has been chosen to
    emphasize the K$^+$\, rows represented by yellow balls}
    \label{crowdion-figure02}
 \end{figure}

Some materials are able to record the passage of charged particles
and are used as radiation
detectors~\cite{crowdion-durrani2008,crowdion-fleischer2011} and
there are minerals that show nuclear tracks that were produced at
some stage during their formation~\cite{crowdion-pricewalker62}.
The mineral mica muscovite has been shown to have recorded the
tracks of muons, positrons and other swift particles with positive
charge while being deep  
underground~\cite{crowdion-russell67a,crowdion-russell67b,crowdion-russell88b}.
The most recent reviews are Refs.~
~\cite{crowdion-russelltracks2015,crowdion-russellcrystal2015}.
The tracks are recorded within the cation layer of potassium ions
which form a two-dimensional hexagonal lattice. There are also
    many tracks along the close packed directions of this lattice that
cannot be produced by charged particles and are attributed to some
vibrational entities called {\em quodons} because of their quasi
one-dimensional
structure~\cite{crowdion-russellcollins95a,crowdion-russellcollins95b,%
crowdion-schlosserrussell94,crowdion-russelleilbeck2011}. Their
existence has also been shown directly with an experiment in which
the energy of alpha particles incident on one side of a mica
specimen was able to eject atoms at the opposite border along the
cation lattice directions~\cite{crowdion-russelleilbeck07}.

Recently, a model with realistic potentials for the dynamics of
potassium ions within the cation layer of mica muscovite has been
introduced~\cite{crowdion-archilla2013,%
crowdion-archilla-springer2014,crowdion-archilla-kosevich2015}.
The authors have considered the available potentials  for the
interaction between atoms and ions. For the interaction between
potassium ions K$^+$\, the electrical potential
  \rcgindex{\myidxeffect{E}!Electrical potential}
   \rcgindex{\myidxeffect{P}!Potential (electrical)}
 was not enough because the passage of the kink brings about very
short distances, for which the ions can no longer be described as
point charges. Therefore, the
 \rcgindex{\myidxeffect{Z}!Ziegler-Biersack-Littmark (ZBL) potential}
  \rcgindex{\myidxeffect{P}!Potential (Ziegler-Biersack-Littmark-ZBL)}
 Ziegler-Biersack-Littmark (ZBL) potential was used~\cite{crowdion-ziegler2008}.
 This potential models the electrical repulsion
by the {\em nuclei} partially screened by the electron cloud. ZBL
potentials have been widely tested and compared to data obtained
in ion bombardment experiments, being therefore the more realistic
ones while using classical mechanics. Quantum calculations could
certainly provide more accuracy but at the cost of much more
complex analytical and numerical calculations. The interaction of
the potassium ions with the lattice was described with empirical
potentials used in molecular dynamics and fitted with
thermodynamic properties, neutron~\cite{crowdion-CollinsAM92} and
infrared spectroscopy~\cite{crowdion-diaz2000} and also validated
for other silicates~\cite{crowdion-gedeon2002}.

Arguably, the most important result in the full system with
substrate was that a supersonic kink
 \rcgindex{\myidxeffect{S}!Supersonic kink}
 \rcgindex{\myidxeffect{K}!Kink (supersonic)}
  was formed with
 specific energy and
velocity~\cite{crowdion-archilla-kosevich2015}. As it involves the
movement of an interstitial atom through the lattice, it will be
called  a (supersonic) {\em crowdion}
 \rcgindex{\myidxeffect{C}!Crowdion}
 in this work as described in
Ref.~\cite{crowdion-kosevich73}. The term will be reserved for
this specific supersonic kink with stable and unique velocity and
not for other kinks. If the lattice was given more energy,
nonlinear waves and later phonons were emitted until the specific
velocity and energy was reached. This characteristic of supersonic
kinks associated with specific values of the velocity have also
been described in
Refs.~\cite{crowdion-savin95,crowdion-zolotaryuk97}, where they
use the terms topological soliton and lattice soliton.

\begin{figure}[b]
\centerline{\includegraphics[width=8cm]{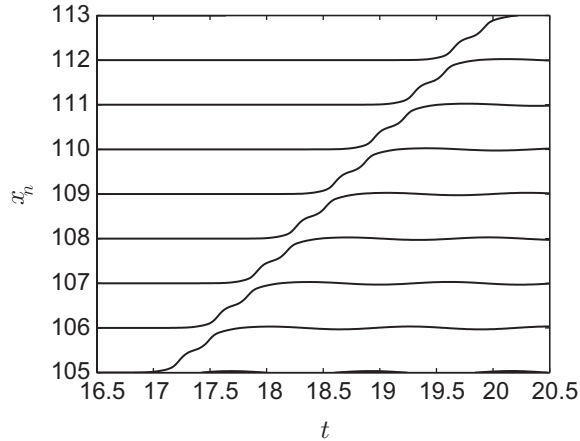}} %
    \caption{Coordinates of the supersonic crowdion or  ultradiscrete kink from numerical simulations.
    It can be observed that only two particles are moving at the same time. Lattice units $a=5.19$\AA\,for
    coordinates and scaled units (0.2\,ps) for time. Also the double kink
    structure can be seen as  will be explained later in the text}
    \label{crowdion-figure03}    
 \end{figure}

  \rcgindex{\myidxeffect{S}!Supersonic crowdion}
    \rcgindex{\myidxeffect{C}!Crowdion (supersonic)}
The supersonic crowdion found in the mica model is extremely
discrete as basically only two ions are moving at the same time,
which will be referred to as the {\em magic mode with sinusoidal
waveform} and corresponds to a phase delay very close to
$q=2\pi/3$~\cite{crowdion-kosevich93,crowdion-kosevich04} as
explained below. In the {\em magic mode}, which was introduced in
the Fermi-Pasta-Ulam lattice to describe both steady-state or
slowly-moving breathers and supersonic kinks
\cite{crowdion-kosevich93}, only two particles are mostly involved
in the motion at the same time. The mode with mode $q=\pi$ is the
limit of discreteness as only one particle is moving at the same
time, and the kink is equivalent to just one particle hitting the
following one with a hard-sphere interaction. We have also called
these kinks {\em ultradiscrete kinks} (UDK).
 \rcgindex{\myidxeffect{U}!Ultradiscrete kink}
 They are also known as kinks with atomic scale localization and
have been described theoretically~\cite{crowdion-friesecke2002}
and observed experimentally in a chain of repelling
magnets~\cite{crowdion-moleron2014}. The energy dissipated by
 the crowdion and its subsequent stability  can be seen
in Fig.~\ref{crowdion-figure01}. Supersonic kinks with a discrete
set of velocities for which there is no radiation have been
described in previous
publications~\cite{crowdion-kosevich73,crowdion-milchev90,%
 crowdion-savin95,crowdion-zolotaryuk97}. They appear in
systems with substrate potential
 \rcgindex{\myidxeffect{S}!Substrate potential}
  \rcgindex{\myidxeffect{P}!Potential (substrate)}
  and nonlinear coupling and can be described as multiple solitons.
In our system due to the extreme discreteness of the kinks there
is only a non-radiating velocity corresponding to a double soliton
as  will be explained in Sect.~\ref{sec:crowdion-crowdion}. See
also Ref.~\cite{crowdion-archilla-kosevich2015}.
Figure~\ref{crowdion-figure03} represents the coordinates of the
potassium ions obtained in a numerical simulation.

The energy of the crowdion is approximately
$E_k=26.2\,\mathrm{eV}$, which is an interesting result because
there are sources of energy in the lattice which can provide it as
it will be explained with more detail in
Sect.~\ref{sec:crowdion-K40}. The most abundant of the unstable
potassium isotopes is $^{40}$K, which can decay by different beta
processes providing recoil energies up to 50\,eV. The crowdion
energy is also smaller than the second ionization energy of
potassium, that is, 31,6\,eV\,\cite{crowdion-lide2010}, which thus
prevents the possibility of inelastic collisions where the kinetic
energy would be lost stopping the propagation of the kink. It is
also larger that the 4-8\,eV needed to eject an
atom~\cite{crowdion-kudriavtsev2005}, an effect that has been
found in an experiment where the transmission of localized energy
along lattice directions with the subsequent ejection of an atom
at the edge of the boundary has been
observed~\cite{crowdion-russelleilbeck07}.
   \rcgindex{\myidxeffect{S}!Sputtering (transmission)}
     \rcgindex{\myidxeffect{T}!Transmission sputtering}

 Another point of interest of the crowdion is that it consists essentially of a charged interstitial $K^+$,
i.e. an excess of an unit of elemental charge, travelling at twice
the speed of sound. Therefore, it is very likely to be recorded,
as positively charged particles leave tracks in mica muscovite.
   \rcgindex{\myidxeffect{C}!Charge of crowdions}
     \rcgindex{\myidxeffect{C}!Crowdions with charge}
   \rcgindex{\myidxeffect{C}!Charge of kinks}
     \rcgindex{\myidxeffect{K}!Kinks with charge}

Are the quodons observed in mica muscovite the crowdions described
in this work? It is not clear, but there are several points in
their favour: a)~They have an energy that can be produced by the
recoil of $^{40}$K; b)~They have enough energy to expel an atom at
the surface; c)~They have stability and seem to travel forever;
d)~They survive to room and higher temperatures; e)~They transport
positive charge that would leave a track in mica muscovite.
 Against them is that
their existence and stability has not been verified in two or
three dimensions. But even if their energy spreads they are likely
to leave some of the other dark marks in mica.

The sketch of this work is as following:
Section~\ref{sec:crowdion-description} describes the system and
potentials. In Sect.~\ref{sec:crowdion-magic} the magic mode is
described with detail and the quantities in the fundamental ansatz
are redefined in a new meaningful way.
Section~\ref{sec:crowdion-crowdion} describes the properties of
the kinks when the substrate potential is introduced and the
supersonic crowdion appears, while
Sect.~\ref{sec:crowdion-phonons} describes the properties of
phonons in a system with a substrate and applies them to analyze
the crowdion's phonon tail. Some interesting results of the
outcome of numerical simulations when excess energy is delivered
and when the system is previously thermalized are presented in
Sect.~\ref{sec:crowdion-simulations}. The recoil energies in the
different decay modes of $^{40}$K and their consequences for the
formation of kinks or other lattice excitations are described in
Sect.~\ref{sec:crowdion-K40}. The work ends with a summary.

\section{Description of the system}
\label{sec:crowdion-description} Mica muscovite is a layered
silicate where a layer of potassium ions is sandwiched between
layers of a complex silicate structure of tetrahedra and
octahedra. This cation layer has a hexagonal structure where rows
of potassium ions can be identified, as seen in
Fig.~\ref{crowdion-figure02}.
        \rcgindex{\myidxeffect{S}!Silicate}
           \rcgindex{\myidxeffect{C}!Cation layer}
           \rcgindex{\myidxeffect{L}!Layer of cations}
As explained with more detail in
Refs.~\cite{crowdion-archilla2013,crowdion-archilla-springer2014,crowdion-archilla-kosevich2015}
we consider an 1D model for a row of K$^+$\, ions. The distance
between ions is $a=5.19$\AA\, which in scaled units will be taken
as the unit of distance. The interaction between ions is described
by two terms, the first one is the electrostatic Coulomb repulsion
             \rcgindex{\myidxeffect{C}!Coulomb potential}
             \rcgindex{\myidxeffect{P}!Potential (Coulomb)}
\begin{equation}
U_C=K_e\frac{\mathrm{e}^2}{r}-K_e\frac{\mathrm{e}^2}{a}
\,,\label{eq:crowdion-coulomb}
\end{equation} where
$K_e$ is the Coulomb constant, $\mathrm{e}$ the elementary unit of
charge and $r=d_n=x_n-x_{n-1}$ is the interatomic distance. The
reference level of energy is taken as the electrostatic energy at
the equilibrium distance $a$.  This value of energy
$K_e\mathrm{e}^2/a=2.7746\,\mathrm{eV}$ is also taken as the unit
of energy in scaled units, and it is useful to remember that it is
approximately $u_E\sim3\,\mathrm{eV}$. The other natural units are
the potassium mass $m_K=39.1$\,amu and therefore the derived unit
of time $\tau=\sqrt{m_K a^3/K_e \E^2}=0.1984\,\mathrm{ps}\simeq
0.2\,\mathrm{ps}$.

\begin{figure}[b]
\centerline{\includegraphics[width=9cm]{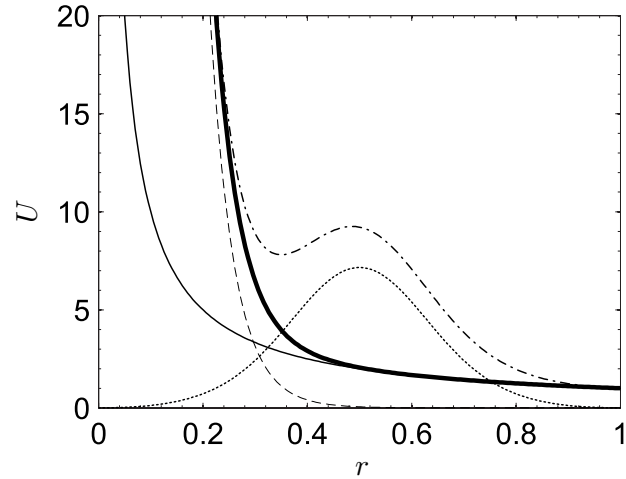}} 
    \caption{Interaction potentials U(r) in scaled units.
Coulomb (---); ZBL ($-\,-$); Coulomb+ZBL (thick --\,); substrate
potential ($\cdot\cdot\cdot$) and the sum of the Coulomb, ZBL and
substrate potentials ($-\cdot-$). The scaled units are 2.77\,eV
and the lattice unit $a = 5.19$\AA\, for $U$ and $r$,
respectively}
   \label{crowdion-figure04} 
 \end{figure}

This system supports propagating kinks of almost any velocity and
energy~\cite{crowdion-archilla2013,%
crowdion-archilla-springer2014,crowdion-archilla-kosevich2015} but
with very small inter-particle distances for which the ions cannot
be described as point particles. The second term for short-range
repulsion is the Ziegler-Biersack-Littmark or ZBL potential, which
corresponds to the electrostatic interaction between nuclei
partially shielded by the electron cloud which is described by an
universal function that has been tested with experiments of ion
bombardment~\cite{crowdion-ziegler2008}. The ZBL potential usually
involves four terms which are effective at different ranges of
energies. For the potassium atoms at energies up to 200\,keV it is
enough to consider a single term given by
 \begin{equation}
U_{\rm ZBL}(r)=\frac{\alpha}{r}\exp(-\frac{r}{\rho})\,,
\label{eq:crowdion-ZBLpotential}
 \end{equation}
with $\alpha=2650.6$\,eV\AA\,  and $\rho=0.29529$\,\AA\, which
correspond to  $\alpha=184.1$ and $\rho=0.0569$ in scaled units,
respectively. No attractive Van der Waals term is considered as it
would be much weaker than the repulsive term. The system with
Coulomb and ZBL potential also support propagating kinks with many
energies but with realistic distances between
particles~\cite{crowdion-archilla-kosevich2015}.

The interaction with the atoms in the lattice above and below the
potassium layer is obtained from an unrelaxed lattice using
empirical potentials used in molecular dynamics and fitted with
thermodynamic and spectroscopic
properties~\cite{crowdion-CollinsAM92,crowdion-gedeon2002} which
are also valid for other silicates. The resulting periodic
potential can be written as a Fourier series for which it is
enough to retain the first five
terms~\cite{crowdion-archilla-kosevich2015}
\begin{eqnarray}
U_s(x)=\sum_{n=0}^4\,U_n\cos(2\pi\,n\,\frac{x}{a})\,.
\end{eqnarray}
The Fourier coefficients are given by
 \begin{eqnarray}
 U_n&=&[6.7902, -9.2920,
3.0512, -0.6387, 0.0891]\,\mathrm{eV}=\nonumber\\
&=&[2.4473, -3.3490, 1.0997, -0.2302, 0.0321]\,,
\label{eq:crowdion-fourier}
\end{eqnarray}
with the latter values given in scaled units. As will be shown
later, the linear spatial frequency for the long wavelength limit
becomes 119\,cm$^{-1}$, that is quite close to the experimental
one of 110\,cm$^{-1}$ obtained with infrared
spectroscopy~\cite{crowdion-diaz2000}. A comparison between the
three potentials can be seen in Fig.~\ref{crowdion-figure04}.

\section{The magic mode revisited}
                 \rcgindex{\myidxeffect{M}!Magic mode}
                 \rcgindex{\myidxeffect{M}!Mode (magic)}
\label{sec:crowdion-magic}  In this section we describe the
fundamental ansatz and the variable involved. We will define the
variables in a proper way, as they are not the same as in plane
waves in spite of their analytic similarity. We will use scaled
units for which the equilibrium interatomic distance is the unity
as described above except where stated otherwise.

\subsection{Basic variables}
Some variables used throughout the study are introduced here, together with their definitions:
\begin{description}[]
\item[Position $x_n$:] It describes the position of the particle
labelled $n$. At equilibrium $x_n=n a$, although the origin of $n$
is arbitrary.
\item[Displacement $u_n$:]  It is the separation of the particle $n$ from
the equilibrium position, that is $u_n=x_n-n a$.
\item[Interatomic distance $d_n$:] It is the distance between two particles
or ions. At equilibrium it is equal to the lattice unit $a$, which
in lattice units is the unity, but it will be written often
explicitly for clarity. It is related with the positions and
displacements as $d_n=x_n-x_{n-1}=u_n-u_{n-1}+a$.
\item[Strain $v_n$:] The increase of $d_n$ with respect to the
equilibrium distance, i.e. $v_n=d_n-a$. It is always negative for
the kinks described in this work. It is related with the
displacements as: $v_n=u_n-u_{n-1}$.
     \rcgindex{\myidxeffect{S}!Strain}
\item[Compression $c_n$:] The decrease of $d_n$ with respect to the
equilibrium distance, i.e. $c_n=a-d_n=-v_n$. It is always positive
for the kinks described in this work. It is related to the
displacements as: $c_n=u_{n-1}-u_n$.
\end{description}
                 \rcgindex{\myidxeffect{S}!Strain}
                              \rcgindex{\myidxeffect{C}!Compression (variable)}

\subsection{Fundamental ansatz} As demonstrated in
Refs.~\cite{crowdion-kosevich93,crowdion-kosevich04} for a large
set of kink solutions of Fermi-Pasta-Ulam systems, the strain
$v_n=u_n-u_{n-1}$ can be approximately described by the {\em
fundamental ansatz with sinusoidal waveform}:
               \rcgindex{\myidxeffect{A}!Ansatz (fundamental)}
               \rcgindex{\myidxeffect{F}!Fundamental ansatz}
               \rcgindex{\myidxeffect{S}!Sinusoidal ansatz}
\begin{equation}
 v_n=-\frac{A}{2} (1+\cos(q (na-V t)) \quad  \textrm{with} \quad  -\pi\leq q(na-V t) <\pi \,
 ,
 \label{eq:crowdion-vn}
\end{equation}
where $q=2\pi/3a$ or $q=2\pi/3$ in scaled units with $a=1$ that we
will usually use. The value of $v_n$ is always negative
representing a compression of the bond. This ansatz describes a
moving profile with velocity V that it is better visualized in the
alternative form $v_n=-A\cos^2(q/2(n-Vt))$. At any given time its
value is zero except for a length $\lambda=2\pi/q$ representing
the number of consecutive bonds compressed.  For a given bond $n$
the value of $v_n$ is zero except for an interval of time
$T=2\pi/(qV)$ representing the time during which the bond is
compressed.  Note that $\lambda$ is not a wavelength as there is
no periodic wave and $T$ is not a period as there is no
periodicity in time.

For convenience we will often use the equivalent expression for
the compressions $c_n=-v_n$:
\begin{equation}
 c_n=\frac{A}{2} (1+\cos(\omega t-q n)) \quad  \textrm{with} \quad  -\pi\leq \omega t - q n<\pi \,,
 \label{eq:crowdion-cn}
\end{equation}
where $\omega=qV$ is the rate of variation of the phase
$\phi(n,t)=\omega t-qn$, i.e., $\omega=\partial \phi(n,t)/\partial
t$ but it is not the frequency as there is no periodicity. This
equation will be used in the next subsection as it is easier to
interpret because $c_n$ is always positive, the phase increases in
time and the bonds compressed later have smaller phase.

\begin{figure}[t]
\begin{center}
\includegraphics[width=11.0cm]{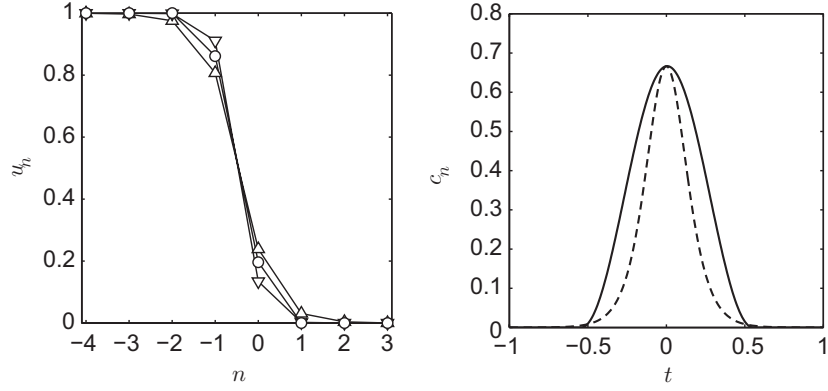}
\caption{({Left} ) Profiles of the displacements $u_n$ for the
sinusoidal magic mode ($\circ$) and the ones given in the original
supersonic crowdions paper by Kosevich and Kovalev
(1973)~\cite{crowdion-kosevich73}. For a quartic  interatomic
potential ($\Delta$): $u_n=(2/\pi)\arctan[\exp(-q(n-V t))]$  and
for a cubic one ($\nabla$): $u_n=[1+\exp(2q(n-Vt))]^{-1}$. ({
Right}) Comparison of the compressions $c_n(t)=u_{n-1}-u_n$ for
the magic mode (--) with the soliton for the continuous  KdV
equation~\cite{crowdion-KdV1895}: $c_n=A\,\mathrm{sech}^2[q(n-V
t)]$ (-\,-). The functions have been rewritten so that the
parameters have the same meaning. The magic mode is between the
two K\&K solutions and it is wider than the KdV one}
    \label{crowdion-figure05}   
    \end{center}
 \end{figure}

From the fundamental ansatz the displacement can be constructed
and  it may be instructive to compare them with other solutions.
They can be seen in Fig.~\ref{crowdion-figure05} for the magic
mode $q=2\pi/3$ compared with the first solutions for supersonic
crowdions~\cite{crowdion-kosevich73}. The compressions
$c_n=u_{n-1}-u_{n}$ have a solitonic form and in the same figure
they are compared with the discretization of the solutions for the
KdV equation, which describes waves in a
canal~\cite{crowdion-KdV1895}, one of the first examples of
solitons.
   \rcgindex{\myidxeffect{K}!KdV soliton}
   \rcgindex{\myidxeffect{K}!Kosevich\ \&\ Kovalev supersonic crowdion}

As we have seen these equations are not as  simple as they seem,
due to the compactness condition for being nonzero. They look like
harmonic waves, but they are not. The quantities in the equation
have to be redefined but they  keep the usual relationships for
harmonic waves. In the following we propose operational
definitions that are convenient but are only approximately
correct, which is also natural at the fundamental ansatz is not
exact either.

\begin{description}[]
\item[\em Velocity $V$:] The average velocity of the kink. This is the
magnitude best defined in numerical simulations and experiments.
\item[\em Phase $\phi(n,t)$:] Trivially, the phase of the bond $n$ is $\phi(n,t)=\omega t -q n$. It
determines when a bond is compressed $-\pi\leq\phi(n,t)<\pi$ and
its state of compression. For example, $\phi(n,t)=0$ is the phase
of the state of maximum compression of the bond $c_n=A$,
$\phi(n,t)=-\pi$ means the beginning of the compression process
and $\phi(n,t)=\pi$ is the end. It is not periodic as a bond is
just compressed once, if for example, $\phi_n=2\pi$ $c_n=0$ and
not $A$.
\item[\em Active:] This term will change depending on the variable we
refer to. For the phase it corresponds to  $\phi(n,t)\in
[-\pi,\pi)$.
      \rcgindex{\myidxeffect{A}!Active bonds}
\item[\em Phase rate  $\omega$:] It is the rate of variation of the
phase with time or $\omega=\partial \phi(n,t)/\partial t=q V$. It
is not the angular frequency as the ansatz is not a periodic
function.
     \rcgindex{\myidxeffect{P}!Phase rate}
\item[\em Compression time $T$:] It is  the  interval
of time for which a bond is compressed or activated,
$T=2\pi/\omega$. The interval of activity starts with zero
compression $c_n=0$ and finishes with the same value. In the
meantime it achieves $c_n=A$, its maximum value.  It also starts
with $\phi=-\pi$ and finishes with $\phi=\pi$. As the numerical
solutions become separate from the fundamental ansatz the
operational definition of $T$ is the value that brings about a
better fit of $v_n$ with the fundamental ansatz.
          \rcgindex{\myidxeffect{C}!Compression time}
\item[\em Phase delay $q$:] It is the
phase difference between two active (compressed)  bonds  $n$ and
$n-1$, that is, $q=\phi(n,t)-\phi(n-1,t)$. Alternatively, it can
be defined as  $q=2 \pi (\delta t/T)=\omega \delta t$, where
$\delta t$ is the time delay between two consecutive active bonds.
\begin{figure}[t]
       \rcgindex{\myidxeffect{P}!Phase delay}
\includegraphics[width=7cm]{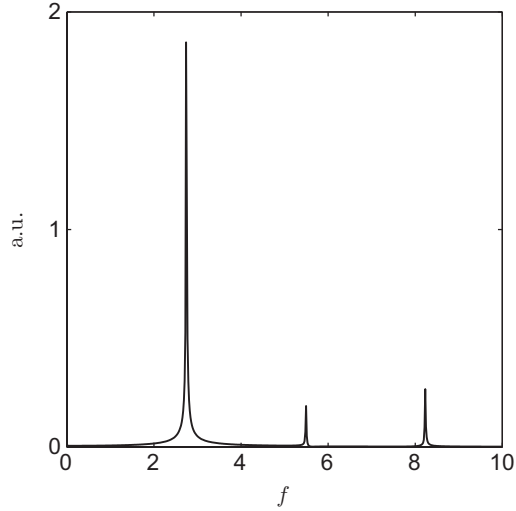} 
\sidecaption[t] \caption{Fourier spectrum of the kinetic energy of
the supersonic crowdion obtained from numerical simulations. It is
measured in a frame that moves with the crowdion in the lattice.
We use arbitrary units for the intensity and scaled units (5\,THz)
for the frequency. The value of the first harmonic is exactly the
characteristic linear frequency $\bar\nu=V_c/a=2.7387$ and
circular frequency $\bar\omega=2\pi\bar\nu\simeq 17.2$, which
corresponds to $\bar \nu\simeq 13.4$\,THz in physical units.}
    \label{crowdion-figure06}    
 \end{figure}

\item[Kink length $\lambda$:] It is the spatial extension of the
kink, very much related with the number of active bonds at a given
time $\lambda/a$ or simply $\lambda$ in scaled units. It is given
by $\lambda=2\pi/q$ and it is also the distance travelled by the
kink during a time interval $T$, i.e., $\lambda=V T$. The usual
relationships also hold, that is, $V= \omega/q$ and
$\lambda=2\pi/q$.
\item[\em Amplitude $A$:]  It is the maximum value of the compression
$c_n$.
        \rcgindex{\myidxeffect{K}!Kink length}
           \rcgindex{\myidxeffect{L}!Length of the kink}
\item[\em Minimum distance $R$:] It is the minimum value of
the interparticle distance $d_n$, that is, $R=a-A$ or $R=1-A$ in
scaled units.
\item[\em Characteristic frequency $\bar{\nu}$:] This is the
inverse of the time $\delta t$ that the kink needs to travel a
distance of a lattice site, i.e. $\bar{\nu}=1/\delta t =V/a$ or
simply $\bar \nu=V$ in scaled units. Note that $\bar
\nu=(\lambda/a)(1/T)$ (and {\em not} $1/T$). As the kink is not
periodic it is the physical frequency at which the compression,
the kinetic or potential energy or other magnitudes change while
the kink travels in a lattice with period $a$.
 An example can be seen in
Fig.~\ref{crowdion-figure06}.
 Their values for the crowdion are therefore $\bar\nu=2.7387$ and
 $\bar{\omega} =2\pi\bar\nu\simeq 17.2$, corresponding to $\bar\nu\simeq 13.4$\,THz.
      \rcgindex{\myidxeffect{C}!Characteristic frequency}
            \rcgindex{\myidxeffect{F}!Frequency (characteristic)}
\end{description}

The equations for the displacements $u_n$ and its derivatives will
be seen in the following subsection.

\subsection{Phasors for the magic mode}
\label{subsec:crowdion-phasors}  The easiest way to visualize the relative
phases and distances of the variables is to consider the rotating
complex vectors or {\em phasors}
 \rcgindex{\myidxeffect{P}!Phasors}
 \rcgindex{\myidxeffect{R}!Rotating complex}
\begin{eqnarray}
\vec{b}_n=\frac{A}{2}\E^{\displaystyle \I\phi(n,t)},\quad
\mathrm{with}\quad \phi(n,t)=\omega t-q n \quad \mathrm{and}\quad
c_n=\frac{A}{2}+\mathrm{Re}(\vec b_n)\,,
\label{eq:crowdion-phasors}
\end{eqnarray}

\begin{figure}[b]
\centerline{\includegraphics[width=10cm]{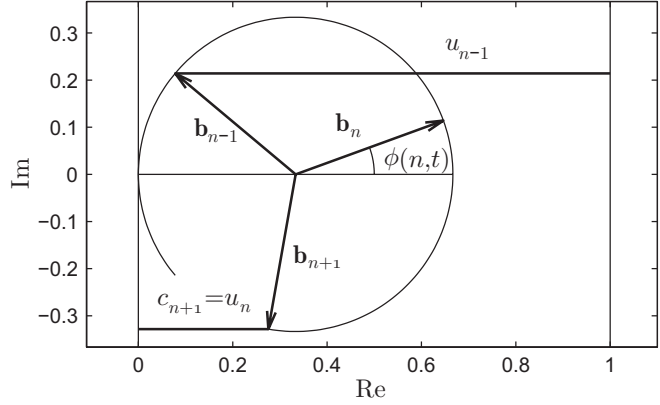}} 
\caption{Visualization of the evolution of the compressions
$c_n=-v_n=u_{n-1}-u_{n}$ for the magic mode $q=2\pi/3$ ($A=2/3$).
Three phasors $\vec{b}_{n-1},\vec{b}_{n},\vec{b}_{n+1}$ centered
at $(0,A/2)$ and rotating anti-clockwise are {\em active} (the
bonds are compressed) at a given time $t$ when $-\pi\leq
\phi(n,t)<\pi$. Their horizontal coordinates give the compression
as can be seen with $c_{n+1}$. The maximum compression A is
achieved for $\phi(n,t)=0$. At $\phi(n,t)=\pi$, $\vec{b}_{n-1}$
will transform into $\vec{b}_{n+2}$ indicating that the bond $n-1$
is no longer compressed while the bond $n+1$ starts its
compression cycle. The displacements are {\em active} while
changing and only two are active at a given time $u_n=c_{n+1}$ and
$u_{n-1}=c_{n}+c_{n+1}=3A/2-c_{n-1}$. For $m>n$, $u_{m}=0$ and for
$m<n-1$, $u_{m}=1$. Also the nonzero velocities are
 $\dot u_n=-\omega\,\mathrm{Im}(\vec{b}_{n+1})$ and $\dot u_{n-1}=\omega
 \,\mathrm{Im}(\vec{b}_{n-1})$. Magnitudes are in lattice units $a=5.19$\,\AA}
\label{crowdion-figure07}
 \end{figure}

\begin{figure}[b]
\centerline{\includegraphics[width=\textwidth]{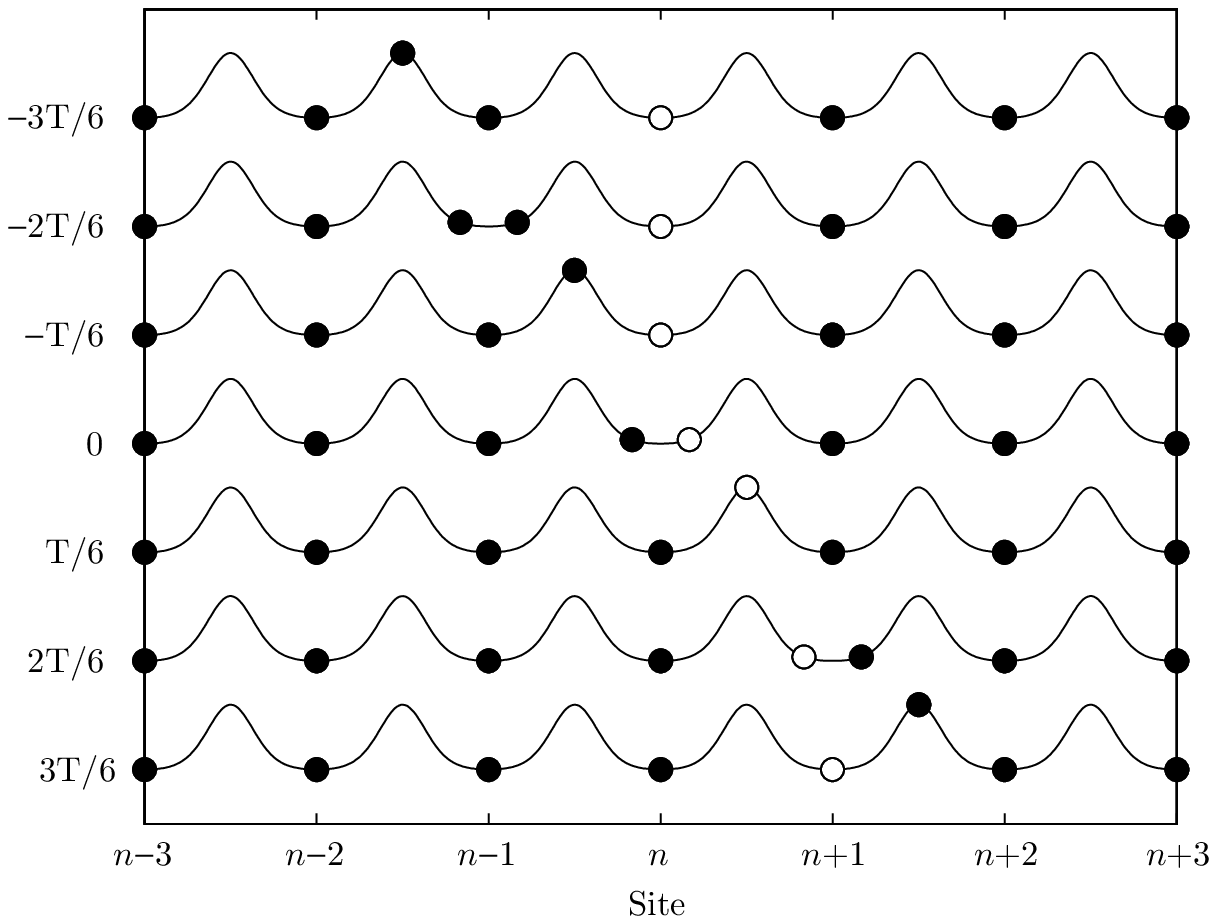}} 
    \vskip-3mm\caption{Magic mode $q=2\pi/3$ for a kink.
    A sketch of the system is shown for a full time of compression $T$
    at time intervals $T/6$. The white particle is labelled $n$,
    therefore its displacement is $u_n$ and the bond at its left
    is also bond $n$ with compression $c_n=-v_n=u_{n-1}-u_n$.
    The origin of time has been taken as the time of maximum
    compression of bond $n$, i.e., $c_n=A$ and $d_n=a-A$.
    During the time interval in the graph the white particle $n$  moves
    from site $n$ to site $n+1$.
    At time $t=-T/2=-3T/6$, the bond $n$ is uncompressed ($c_n=0$,
    $d_n=1$) and again becomes uncompressed at $t=3T/6=T/2$.
    Note that during the first two $T/6$ intervals, although the bond $d_n$ is
    changing, there is no appreciable displacement $u_n$. Note also, that the
    compressed structure at $t=-3T/6$  between sites $n-2$, $n-1$
    has moved at $t=3 T/6$ to sites $n+1$, $n+2$, i.e., the
    kink has moved three sites or the length of the kink $\lambda=V_c T$, while the
    white particle $n$ has moved a single site. Therefore, the
    average velocity of a particle in a time $T$  is $\langle V_p\rangle=1/T=V_c/3$.
    The average velocity of a particle for the following four $T/6$ intervals, when
    it is actually moving, is $\langle V_p\rangle'=1/(2T/3)=V_c/2$}
    \label{crowdion-figure08}
 \end{figure}

There is an important difference with the usual concept of phasors
and it is that the circle is not periodic. The only phase interval
where the phasors exists is  $-\pi\leq\phi(n,t)<\pi$. If
$\phi(n,t)<-\pi$ the phasor $\vec b_n$ has not yet come into
existence and when $\phi(n,t)>\pi$,  $\vec b_n$ has disappeared.
Therefore, for $q=2\pi/3$ at a given time there are three phasors
in the unit circle as shown in Fig.~\ref{crowdion-figure07}. The
three phasors have their origin at $(A/2,0)$ and rotate
anti-clockwise with angular speed $\omega$ while the time $t$
increases, let us denote them $\vec{b}_{n-1}$, $\vec{b}_n$,
$\vec{b}_{n+1}$. In the following $n$ has to be understood as the
index of the inner bond of the three compressed ones or the index
of the intermediate phasor, that is $-\pi/3 \leq\phi(n,t)<\pi/3$.
If we denote as $t_n=n/V$ the time for which $\phi(n,t_n)=0$, then
$-T/6\leq t-t_n <T/6$. This is not a restriction as there is
always a bond central to the three compressed ones.

The phasor $\vec b_{n+1}$ is behind $\vec b_{n}$ ny an angle $q$
and so on for a kink travelling to increasing $n$ number. Note
that $\vec{b}_{n-1}+\vec{b}_n+\vec{b}_{n+1}=0$.

Therefore, the particles first reached by the kink have larger
phase $\phi$. The angle $\phi=\pi$ is the angle for change of
number, that is, when $\vec b_{n-1}$ reaches that position it
disappears from the circle and ceases to be active, indicating
that the bond $n-1$ is no longer compressed. At the same time, a
new phasor $\vec b_{n+2}$ appears at $\phi=-\pi$, indicating that
a new bond has started to be compressed or becomes {\em active},
after a time $T$ it will in turn become inactive. As shown in
Fig.~\ref{crowdion-figure07}, the horizontal distance to the
vertical straight line through the origin is the compression
$c_n=A/2+\mathrm{Re}({\vec b}_n)$.

Let us now consider the displacements $u_n$, using
$c_n=u_{n-1}-u_n$ or $u_{n-1}=u_n+c_n$. The  particles not yet
reached by the kink have zero displacement and the first nonzero
compression is $c_{n+1}$. Therefore $u_{n}=c_{n+1}$ and
$u_{n-1}=u_n+c_n=c_n+c_{n+1}=A+\mathrm{Re}(\vec{b}_n+\vec{b}_{n+1})=A+\mathrm{Re}(-\vec{b}_{n-1})=
 3 A/2-c_{n-1}$ as represented in Fig.~\ref{crowdion-figure07}. To
 summarize
\begin{eqnarray}
u_{n+1}&=&0\\\nonumber
 u_n&=&c_{n+1}=\frac{A}{2}+\mathrm{Re}(\vec{b}_{n+1})=\frac{A}{2}+\frac{A}{2}\cos(\omega t-q)  \nonumber\\
 u_{n-1}&=&\frac{\displaystyle 3A}{\displaystyle 2}-c_{n-1}= A-\mathrm{Re}(\vec{b}_{n-1})=A-\frac{A}{2}\cos(\omega
 t+q)\,.
\end{eqnarray}
These equations are valid for $t=0$ chosen as the time for which
the central bond $n$ is most compressed $c_n=A$ and remains
central, $-\pi/3\leq \phi(n,t)<\pi/3$ and $-T/6\leq t<T/6$.
 The following displacement
$u_{n-2}=c_{n-1}+c_n+c_{n+1}=3A/2=1$ and equally $u_m=1$ for
$m\leq {n-1}$, that is, for the particles that have been left
displaced by a lattice unit after the passage of the kink.

 The particle velocities
 $\dot u_m=\partial u_m/\partial t $ can also be calculated  and visualized easily using
 $ \vec{\dot b}_m=\I \omega \vec{b}_m $ and therefore
 $\mathrm{Re}(\vec{\dot b}_m)=-\omega \mathrm{Im}(\vec{b}_m)$
\begin{eqnarray}
\dot u_n&=&\dot c_{n+1}= -\omega\,\mathrm{Im}(\vec{b}_{n+1})=-\omega \frac{A}{2}\sin(\omega t-q)  \nonumber\\
 \dot u_{n-1}&=&-\dot c_{n-1}=\omega \,\mathrm{Im}(\vec{b}_{n-1})=\omega\frac{A}{2}\sin(\omega t+q)\,.
\label{eq:crowdion-dotunmagic}
\end{eqnarray}
For any other $m$, $\dot u_m=0$.

For other integer values of $\lambda=2\pi/q$, there are $\lambda$
active phasors and for non integer values, the number of active
 \rcgindex{\myidxeffect{A}!Active phasors}
  \rcgindex{\myidxeffect{P}!Phasors (active)}
phasors changes between the two integers below and above
$\lambda$. However, in this work we will concentrate on the {\em
magic} mode $q=2\pi/3$ as it is very close to the crowdion found
in the simulations.

In this way it is easy to construct the evolution of the particles
during  the compression time $T$ as can be seen in
Fig.~\ref{crowdion-figure08} for six times between $-T/2$ to
$T/2$. In this time the crowdion advances a length $\lambda=3$,
that is, three lattice units, but a single particle just travels a
single lattice unit. Therefore the average velocity of a particle
$<V_p>$ is three times smaller than the crowdion velocity $V_c$.
It is worth mentioning that Fig.~\ref{crowdion-figure08} also
shows that only the two particles participating in the kink motion
are mostly involved  in the motion at the same time,
 as the fundamental ansatz with sinusoidal waveform, Eq.~\ref{eq:crowdion-vn}  with
$q=2\pi/3$  predicts
\cite{crowdion-kosevich93,crowdion-kosevich04}.

\section{Kinks with substrate potential: the crowdion}
\label{sec:crowdion-crowdion}
\begin{figure}[t]
\centerline{\includegraphics[width=8cm]{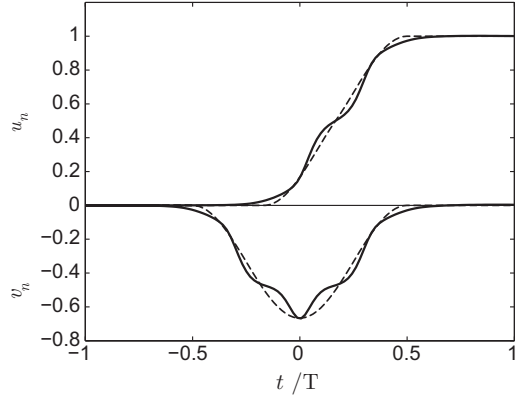}} 
\caption{Comparison of the ultradiscrete kink defined with the
fundamental ansatz in Eq.~\ref{eq:crowdion-vn} with $A=2/3$ and
$q=2\pi/3$ with the ultradiscrete kink with fixed velocity
obtained in the simulations dubbed {\em crowdion} in this work.
Dashed lines: ansatz, continuous lines: crowdion. The
displacements $u_n$ correspond to the upper curves while the
strains $v_n=u_n-u_{n-1}$ correspond to the lower curve. The kink
transforms into a double kink
  \rcgindex{\myidxeffect{D}!Double kink}
    \rcgindex{\myidxeffect{K}!Kink (double)}
 because the displacement between two equilibrium sites is
divided by the nonequilibrium position at the top of the potential
well as can be seen in Fig.~\ref{crowdion-figure08}. The
magnitudes $u_n$ and $v_n$ are given in lattice units $a=5.19$\AA.
The compression time is given by $T=1.095$ or 0.22\,ps in physical
units}
    \label{crowdion-figure09}
 \end{figure}

The introduction of a substrate potential also modifies
substantially the behaviour of the particles in the kink. The
phase $\phi(n,t)$ is still very useful for the interpretation of
the movement of the particles. The crowdion, of ultradiscrete kink
of fixed velocity and energy that appears in the simulations
corresponds basically to the {\em magic mode} but with some
differences. Considering the white ball in
Fig.~\ref{crowdion-figure08} and denoting it by $n$, it basically
does not move from $t\in [-3T/6,-T/6]$ as the Coulomb repulsion
from particle $n-1$ is weak. For times close to $t=0$ when the
strong ZBL potential acts, it receives most of its momentum which
it will transfer in due course to the following particle $n+1$.
However, in between, it will have to overcome the barrier of the
potential, experiencing a deceleration and afterwards an
acceleration while going downhill. Eventually the acceleration
becomes negative as it experiences the ZBL repulsion from the
particle $n+1$ ahead. The ascending and descending of the
potential barrier by the particle produces a remarkable change in
the particle displacement $u_n$ and strain $v_n=u_n-u_{n+1}$ as
shown in Figs.~\ref{crowdion-figure09}. The kink has been
converted into a double kink: the first kink corresponds to the
translation of a particle from the well bottom to the top of the
nearest potential barrier and the second kink to the subsequent
displacement to the following well bottom.

We would also like to mention in connection with
Fig.~\ref{crowdion-figure09}, that the fundamental ansatz with
sinusoidal waveform, Eq.~(5) for $q=2\pi/3$, and corresponding
dashed lines in these figures give much better agreement with the
simulations of supersonic kink motion in the Fermi-Pasta-Ulam
lattice without substrate~\cite{crowdion-kosevich04}. The
deviation from the ansatz prediction in
Fig.~\ref{crowdion-figure09} is caused only by the presence of the
substrate because the ansatz was originally proposed for the
translationally-invariant Fermi-Pasta-Ulam
lattice~\cite{crowdion-kosevich93,crowdion-kosevich04}.
\begin{figure}[t]
\begin{center}
\includegraphics[width=\textwidth]{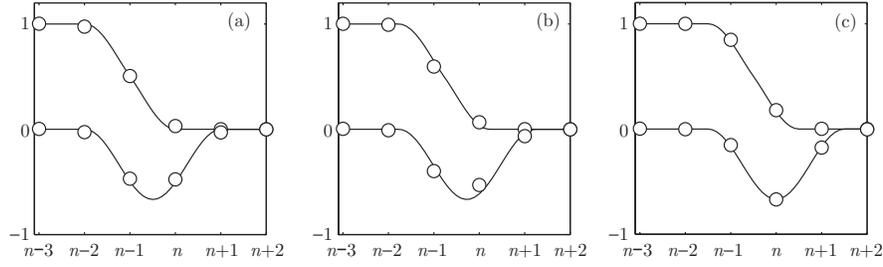}
\caption{Three plots at different times (a) $t\simeq -T/6$ (b)
$t\simeq -0.5T/6$ (c) $t\simeq 0$. They show the profile of the
displacements $u_n$ (upper curves) and strains $v_n=u_n-u_{n-1}$
(lower curves) with respect to the particle and bond index $n$.
The continuous lines represent the theoretical ansatz
Eq.~(\ref{eq:crowdion-vn}) and the circles represent the points
corresponding to the numerical simulation of the crowdion. Time
$t=0$ corresponds to the maximum compression of bond $n$. The
variables $u_n$ and $v_n$ are given in lattice units
$a=5.19$\,\AA. Every $T/6$ the theoretical and numerical solutions
becomes almost identical as can also be seen in
Fig.~\ref{crowdion-figure09}. Subfigure (b) represents the maximum
separation from the theoretical curves}
    \label{crowdion-figure10}    
\end{center}
 \end{figure}

The separation from the ideal functions of the ansatz can also be
seen in Fig.~\ref{crowdion-figure10} where the displacements are
shown at a given time. It can be observed that the deviation from
the magic mode are important qualitatively but not so much
quantitatively. A more significant difference appears in the
velocities which are represented in
Ref.~\cite{crowdion-archilla-kosevich2015} but can also be seen
easily in the slope of Fig.~\ref{crowdion-figure09}. According to
Eq.~(\ref{eq:crowdion-dotunmagic}) the maximum particle velocity
using the ansatz is $\omega A/2=1.91$ or 5\,km/s, while for the
observed one for the crowdion it is 2.9 in scaled units or
7.6\,km/s attained when the particle is going uphill or downhill.
The minimum particle velocity is achieved at the top of the
barrier.

\section{Phonons and crowdions}
\label{sec:crowdion-phonons}

The introduction of the substrate potential brings about
significant changes in the system, not only for the kinks but also
for the phonon spectrum. We first review the properties of phonons
in a system with substrate potential and then use them to analyze
the phonon tail of the crowdion.

\subsection{Phonons in presence of a substrate potential}
 \label{subsec:crowdion-phonsubstrate}
    \rcgindex{\myidxeffect{S}!Substrate potential}
        \rcgindex{\myidxeffect{P}!Phonons with substrate potential}
                \rcgindex{\myidxeffect{D}!Dispersion relation with substrate}
    The dynamical equations for small perturbations become
\begin{equation}
\ddot u_n=-\omega_\mathrm{0}^2 u_n
+c_s^2(u_{n+1}+u_{n-1}-2\,u_n)\, , \label{eq:crowdion-dynphonon}
\end{equation}
with $c_s=\sqrt{2}$. The linearization of the coupling terms come
only from the Coulomb one. The ZBL potential does not appear
because it is zero for small oscillations. The substrate potential
has been reduced to a harmonic one expanding the sinusoidal
functions. The value of $\omega_\mathrm{0}$ is obtained using the
values of the Fourier coefficients of the substrate potential in
Eq.~(\ref{eq:crowdion-fourier})
\begin{equation}
\omega_\mathrm{0}^2=-\sum_{m=1}^4(2\pi m)^2 U_m\,.
\end{equation}
The resulting numerical value is $\omega_0=4.48$ in scaled units,
corresponding to 3.6\,THz or 119\,cm$^{-1}$. The coefficient
 $c_s=\sqrt{2}$ or  3.7\,km/s in physical
units is the speed of sound in the system without
 substrate.

\begin{figure}[t]
\begin{center}
\includegraphics[width=12cm]{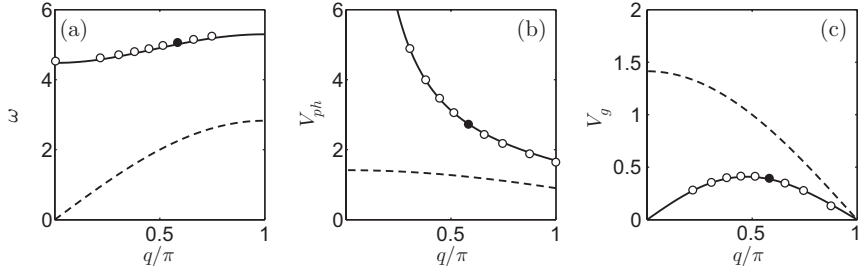}
\caption{(a) Dispersion relation, (b) phase velocity and (c) group
velocity. The three plots are for longitudinal phonons in a
potassium row for the system without substrate (dashed line) and
with substrate (continuous line). Scaled units are approximately
5\,THz for frequency and 2.6\,km/s for velocities. White circles
are measurements from different numerical simulations. The black
circles are the theoretical values for the phonon tail obtained by
making the phonon phase velocity equal to the crowdion velocity}
    \label{crowdion-figure11} 
\end{center}
 \end{figure}

Substitution of $u_n=\exp(\I(qn-\omega t))$ leads to
\begin{equation}
-\omega^2=-\omega_\mathrm{0}^2 +c_s^2( \E^{\displaystyle \I q }+
\E^{\displaystyle -\I q }-2)\,.
\end{equation}
          \rcgindex{\myidxeffect{G}!Group velocity}
                    \rcgindex{\myidxeffect{V}!Velocity (group)}
From this equation it is easy to obtain the phonon spectrum, the
phonon velocities and the group phonon velocities. They are given
by
\begin{eqnarray}
\omega^2&=&\omega_0^2+4\,c_s^2\sin^2(\frac{q}{2})\quad ;\quad V_\textrm{phase}=\frac{\omega }{q} \nonumber\\
 V_g&=&\frac{\D \omega}{\D q}=\frac{c_s^2\sin{q}}{\sqrt{\omega_0^2+4\,c_s^2\sin^2(\frac{\displaystyle q}{\displaystyle 2}})}\,.
 \label{eq:crowdion-wqsubs}
\end{eqnarray}
           \rcgindex{\myidxeffect{P}!Phase velocity}
                      \rcgindex{\myidxeffect{V}!Velocity (phase)}
The corresponding equations for the system without substrate are
identical with $\omega_0=0$. In this case  $c_s$ is both the phase
and group velocity in the long-wavelength limit ($q\rightarrow
0$).

For the system with substrate $\omega_\mathrm{0}$ is the lowest
phonon frequency, corresponding to the long wavelength limit
($q\rightarrow 0$). This can be seen in
Fig.~\ref{crowdion-figure11} where the dispersion relation, the
phase and the group velocities are shown. Note the main changes
produced by the introduction of the substrate potential: (a) the
phonon spectrum becomes optical, i.e., bounded from below, (b) the
phase velocity diverges when $q\rightarrow 0$, and (c) the group
velocity becomes zero both at $q=0$ and $q=\pi$ and has a maximum
close to $q=\pi/2$ but with a much lower velocity.

The value of the wavevector $q$ corresponding to the maximum group
velocity can be calculated as it corresponds to $\D V_g/\D q=0$.
Equivalently it corresponds to the maximum of the function
\begin{equation}
f(q)=\frac{V_g^2}{c^4}=\frac{\sin^2(q)}{\omega_\mathrm{0}^2+2
c_s^2 (1-\cos(q))}\,,
\end{equation}
where we have used that $2\sin^2(q/2)=1-\cos(q)$.
 Then
\begin{equation}
\frac{\D f(q)}{\D q}=\frac{2\sin(q)\cos(q)[\omega_\mathrm{0}^2+2
c_s^2-2 c_s^2 \cos(q)] -\sin^2(q)[2 c_s^2
\sin(q)]}{\omega_\mathrm{0}^2+2 c_s^2-2 c_s^2 \cos(q)}\, .
\end{equation}
Making the numerator equal to zero, we obtain:
\begin{equation}
(\omega_\mathrm{0}^2+2 c_s^2)\cos(q)-2 c_s^2
\cos^2(q)-c_s^2\sin^2(q)=0\,,
\end{equation}
which leads to a second order equation in $\cos(q)$
\begin{equation}
c_s^2\cos^2(q)-(\omega_\mathrm{0}^2+2 c_s^2)\cos(q)+ c_s^2=0\,,
\end{equation}
with solution
\begin{equation}
\cos(q)=\frac{\omega_\mathrm{0}^2+2c_s^2 \pm
\sqrt{\omega_\mathrm{0}^4+4 \omega_\mathrm{0}^2c_s^2}}{2 c_s^2}\,
.
\end{equation}
For the values in the present system, only the minus sign gives a
real value of $q=1.4870$\,rad corresponding to a wavelength
$\lambda=4.2253$, and maximum group velocity $V_{g,M}=0.4091$.

\subsection{Crowdion phonon tail}
             \rcgindex{\myidxeffect{P}!Phonon tail}
                        \rcgindex{\myidxeffect{T}!Tail (phonon)}
\begin{figure}[t]
\begin{center}
\includegraphics[width=12cm]{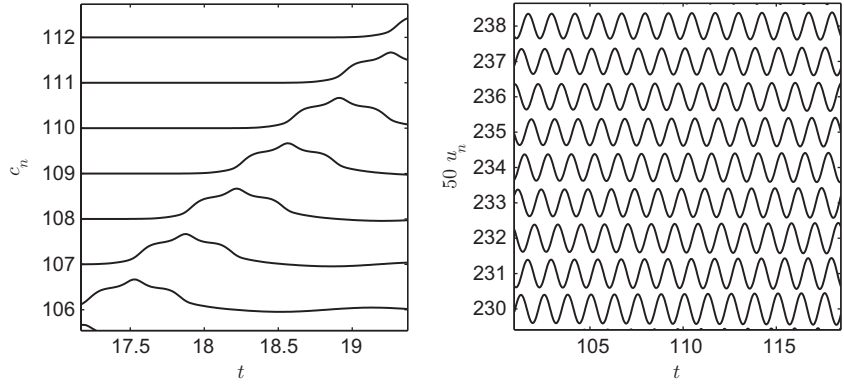}
    \caption{({Left} ) Plot of $c_n+n$ where
    the double soliton structure, period and other crowdion parameters can be appreciated. ({Right})
     Phonon tail amplified 50 times. It is a perfect plane wave with parameters
     with the same velocity of the crowdion $V=V_c$ and similar
    parameters although not identical $T\gtrsim T_c,
    q \gtrsim
     q_c,  \lambda \gtrsim \lambda_c$. Note that this
     parameters are not well defined for the crowdion and only
     approximate}
    \label{crowdion-figure12}    
    \end{center}
 \end{figure}

When the kink is produced, its amplitude diminishes towards the
crowdion's one in an asymptotic way. Therefore after some time,
              \rcgindex{\myidxeffect{A}!Asimptotic}
 the nonlinear waves are no longer produced but there is always a
linear vibration left behind although with decreasingly smaller
amplitude. This is why the crowdion continues propagating. The
tail is a plane wave and as such does not transport energy, but
theoretically could be measured to detect crowdion properties. We
will call it the {\em phonon tail}. Note that the velocity to
describe these plane waves is the phase velocity and which in this
case is
 unbounded. The crowdion is moving at speed $V_c$ and
leaves at each site some small perturbation exactly at the same
estate at times separated by $\delta t=1/Vc$. In other words, the
phase velocity of the phonon tail $V$ is the same as the velocity
of the crowdion $V_c$ .
\begin{equation}
V_\textrm{phase} =V_c=2.7387 \quad \textrm{(7.2\,km/s)}\,\quad
(\textrm{Phonon tail})\,. \label{eq:crowdion-vtail}
\end{equation}
The wave number of the tail can be obtained from the equation
$V_\textrm{phase}=V_c=w/q=[\omega_\mathrm{0}^2+4\sin^2(q/2)]^{1/2}/q$,
which can be solved numerically or graphically from
Fig.~\ref{crowdion-figure11}(b). The result is
$q=1.8290=0.5822\pi$ and therefore $\omega=q V_c=5.00$,
$T=2\pi/\omega=1.2544$ and $\lambda=2\pi/q=3.44$. So the
parameters are very close to the $\omega_c$ , $T_c$ and
$\lambda_c$ of the crowdion. In some sense, they can be considered
as the actual parameters of the crowdion as they can be measured.
Note that these parameters, as $\lambda_c$ are not well defined as
they depend on the algorithm used to fit the numerical solutions.
Figure~\ref{crowdion-figure12} represents a picture of $c_n$ and a
view of the phonon tail for $u_n$, similar to $c_n$ where the
perfect plane wave and its parameters can be appreciated.

\section{Excess energy and  thermalized medium}
\label{sec:crowdion-simulations}
\begin{figure}[tbp]
\begin{center}
  \includegraphics[width=5.5cm]{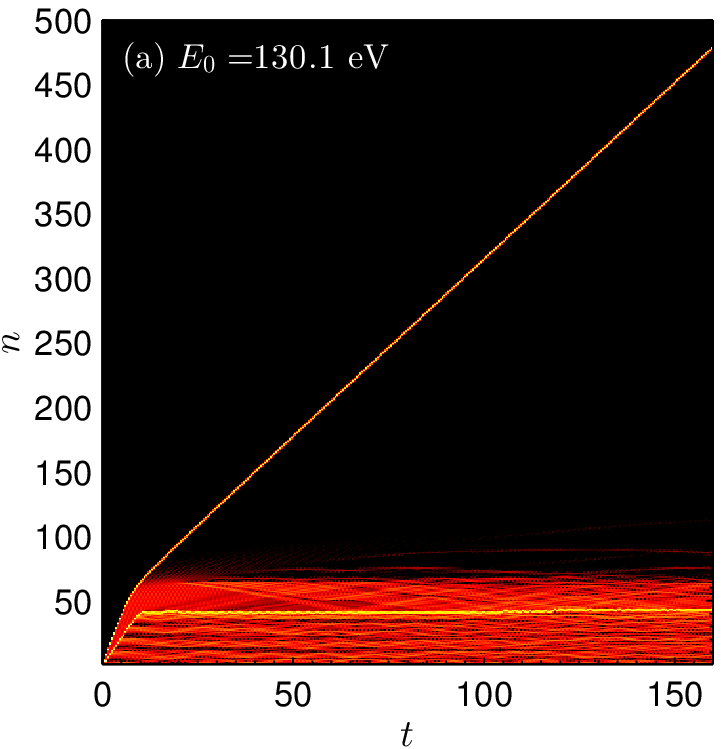}
  \includegraphics[width=5.5cm]{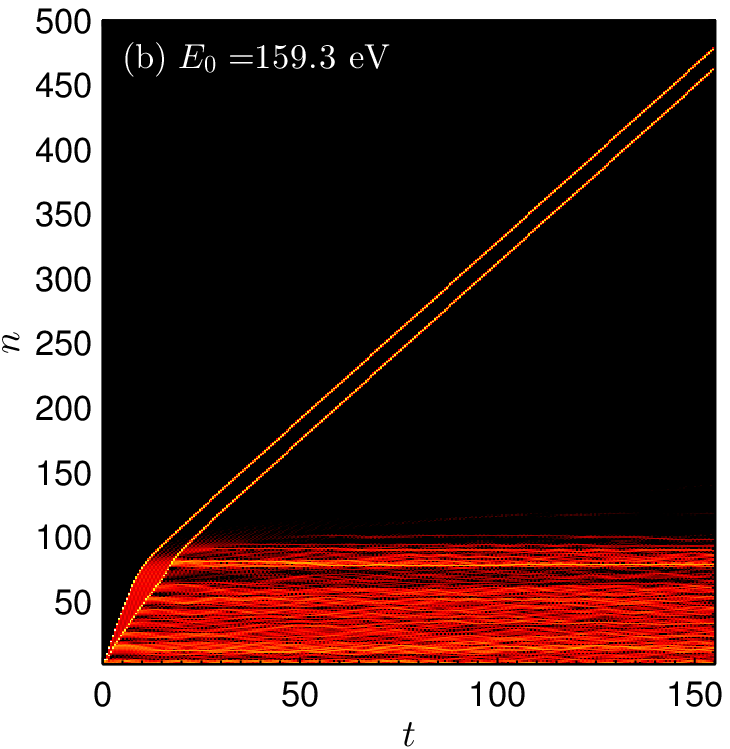}
  \includegraphics[width=5.5cm]{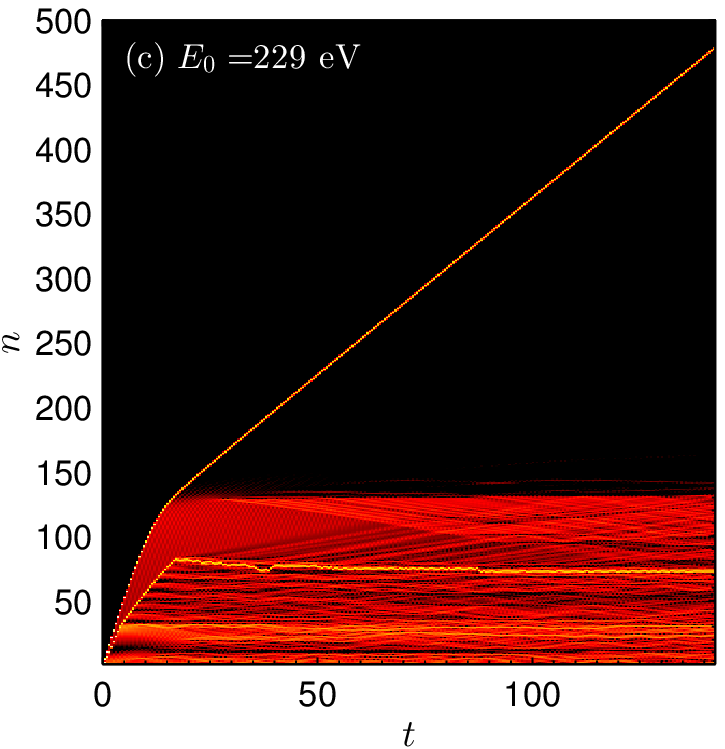}
  \includegraphics[width=5.5cm]{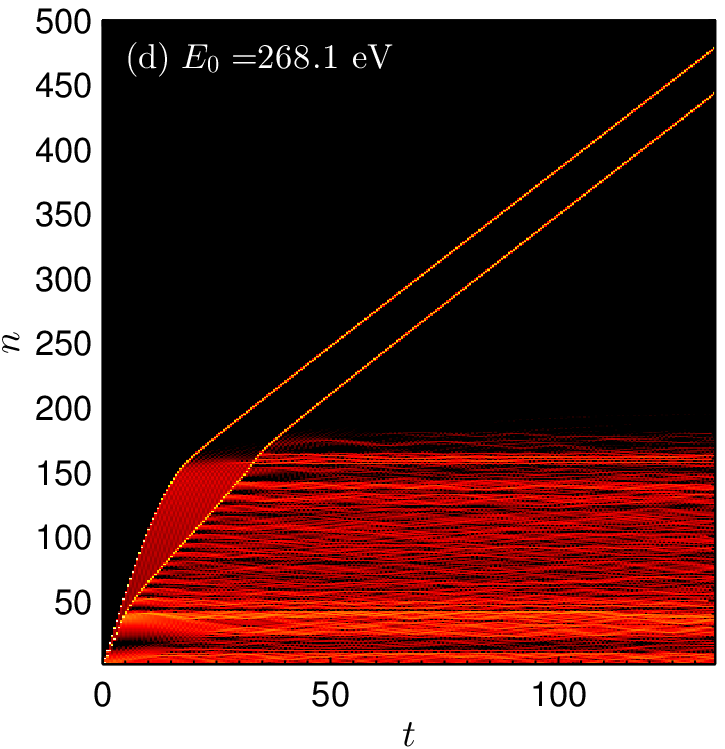}
  \includegraphics[width=5.5cm]{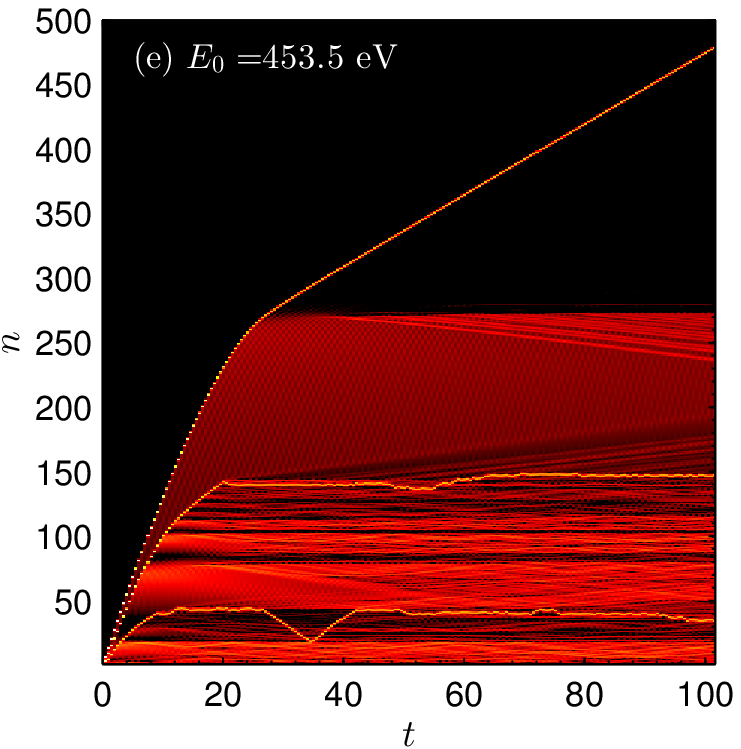}
  \includegraphics[width=5.5cm]{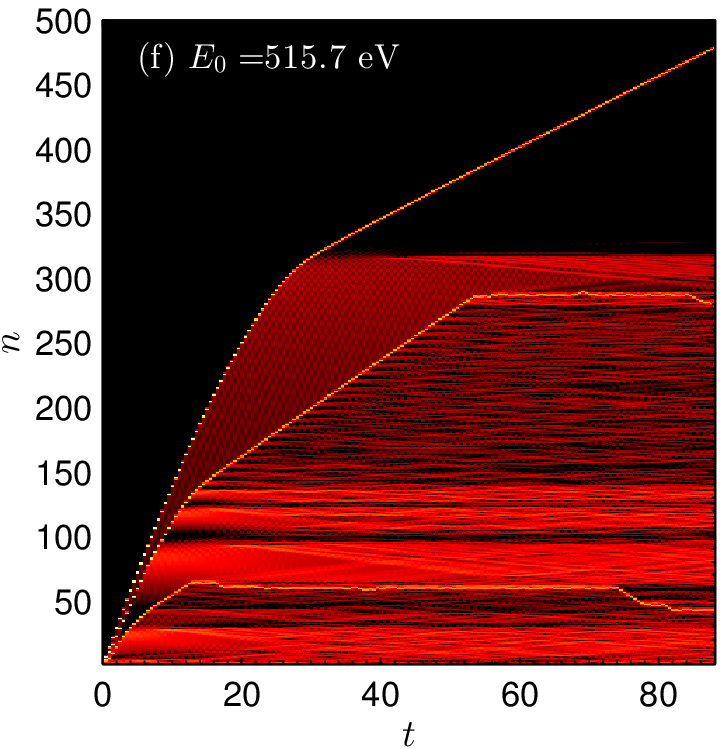}
\caption{(Color) Particle energy plots of several examples of
crowdion formation in arbitrary units of $\simeq 3$\,eV.  Initial
energy increases form (a) to (f). Many features can  be observed,
among them the specific velocity of the crowdion $V_c$, the
formation of nonlinear waves and phonons, the formation of two
crowdions and the survival of the crowdion in the severely
perturbed media for hundreds of sites}
\label{fig:crowdion-figure13}    
\end{center}
 \end{figure}

In this section we present the results of different simulations to
show the capacity of the crowdions to survive a perturbed
environment when larger energy is initially delivered and, second,
the behaviour of the crowdions with temperature.
             \rcgindex{\myidxeffect{E}!Excess energy}

\subsection{Excess energy}
\label{subsec:crowdion-excess} We present some examples of
simulations when the lattice is given more energy than the
26.2\,eV needed to produce the supersonic crowdion. The energies
range from 130\,eV to 520\,eV. They are represented in
Fig.~\ref{fig:crowdion-figure13}. In (a) a single crowdion is
formed after nonlinear waves are emitted. In (b) two crowdions are
formed leaving behind an stationary linear wave. Note how the
second crowdion survives to the tail of the first and the common
velocity $V_c$ of both. In (c) the excessive energy destroys the
second crowdion which transforms into a highly localized nonlinear
stationary wave. In (d) the second crowdion survives again, while
in (e) it is again destroyed. Extensive phonon radiation and
wandering kinks can be seen in the latter figure. In (f) a second
crowdion survives for 150 sites in a highly perturbed media but it
is finally pinned down.

\subsection{Thermalized medium}
            \rcgindex{\myidxeffect{T}!Thermalized medium}
 \label{subsec:crowdion-thermal}
\begin{figure}[p]
\begin{center}
\includegraphics[height=14.4cm]{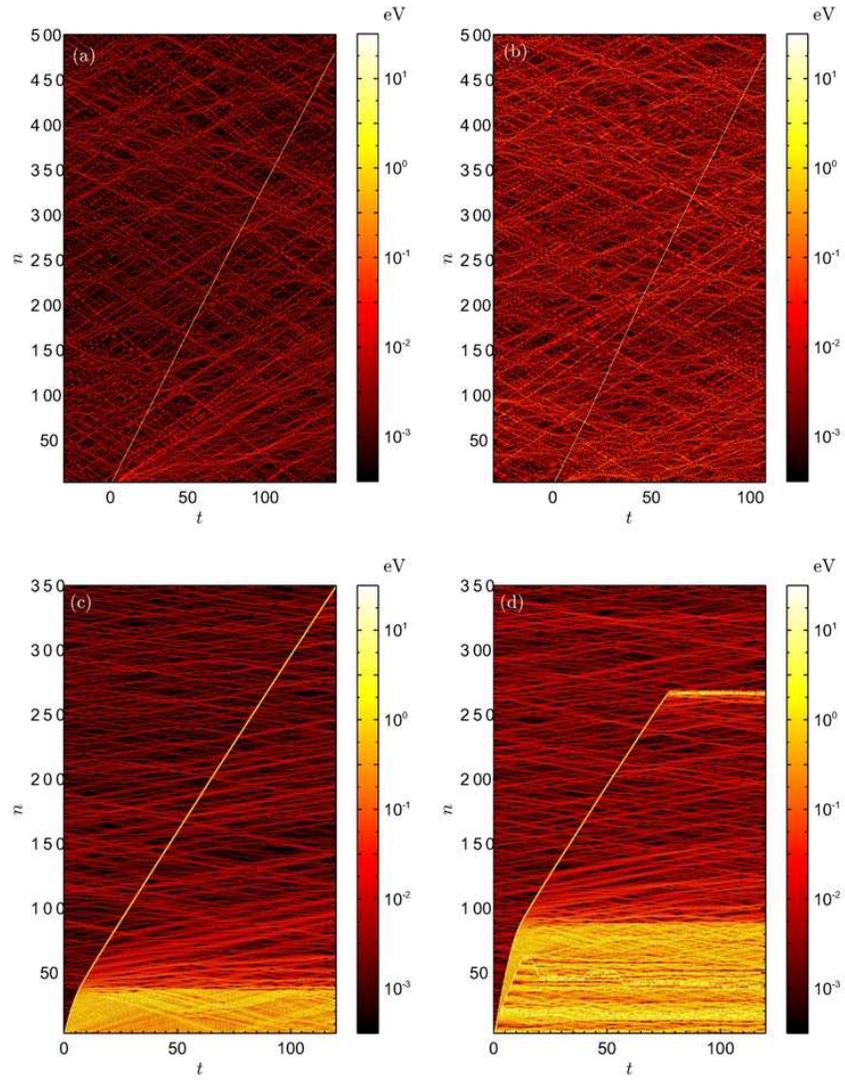}
\caption{(Color) Particle energy plots of two crowdions travelling
in a previously thermalized medium at (a,c)~300\,K and
(b,d)~1000\,K, top without and bottom with substrate potential.
Color bars are in $10\log _{10}(E)$ units}
    \label{crowdion-figure14}   
\end{center}
 \end{figure}

 An
interesting question is whether the crowdion can travel trough a
previously thermalized medium. This is not only a question of
general interest but particulary important for mica muscovite. As
it has been calculated in Ref.~{\em Tracks in mica: 50 years
later}~\cite{crowdion-russelltracks2015}, the recording process of
tracks happens a few kilometers underground under large pressure
and temperatures of 700-1000\,K. Although much more work is
necessary, the answer is positive. For comparison
Fig.~\ref{crowdion-figure14}~(a-b) shows two simulations at 300\,K
and 1000\,K in the system without substrate potential where the
kink survives over hundreds of lattice sites. It is not really
surprising as, if we compare the energy of the crowdion 26.2\,eV
with the mean thermal energy of a particle $k_{B} T$, the crowdion
energy is 1000 and 300 times larger at 300\,K and 1000\,K,
respectively.

In the case of including the substrate potential, as shown in
Fig.~\ref{crowdion-figure14}~(c-d) for 300\,K and 1000\,K
respectively, the crowdion  can also travel for hundreds of sites
of the previously thermalized media. As it was studied in
Ref.~\cite{crowdion-archilla-kosevich2015}, the crowdion always
has finite kinetic energy, but the final total energy of the kink,
$E_k$, is always of the order of magnitude of the Peierls-Nabarro
(PN) barrier. The equivalent kinetic energy equivalent for the
thermalized media is 0.005 (0.013\,eV) at 300\,K and 0.016
(0.043\,eV) at 1000\,K in normalized and physical units . These
values are far below  the energy difference between the PN barrier
and the kink energy. However, in some simulations, for
temperatures of 1000\,K the thermalization is not completely
achieved due to appearance of nonlinear waves instead of phonons.
Therefore, localized peaks of the background vibrations can
interfere with the crowdion where, in some cases, it can be
trapped leading to a highly localized nonlinear stationary
perturbation. Figure~\ref{crowdion-figure14}~(d) shows and example
of this situation, where the crowdion is eventually trapped
forming an interstitial defect.

Thermal effects discussed in this section lead to different
survival path lengths of the crowdions. If the hypothesis of
crowdions propagating in mica muscovite is correct,  they might be
related with some of the tracks observed in the mineral. Other
feature of the presented simulations worth remarking on is that
the high equivalent temperature of the nonlinear tail radiation of
the crowdion is likely to favour a change of structure and the
formation of tracks.

\section{Recoil energy of $^{40}\mathrm{K}$}
\label{sec:crowdion-K40}
             \rcgindex{\myidxeffect{R}!Recoil energy of $^{40}$K}
            \rcgindex{\myidxeffect{P}!Phase velocity}
           \rcgindex{\myidxeffect{D}!Decay $^{40}$K}
                      \rcgindex{\myidxeffect{K}!$^{40}$K}
If the hypothesis of quodons being vibrational entities of ions of
potassium is correct, the most likely source of energy is the
recoil from $^{40}\mathrm{K}$ because a)~the energy will be given
directly to the potassium ion K$^+$\,, b)~the relative abundance
and decay frequency of $^{40}\mathrm{K}$, and c)~because of the
energies involved as explained below.

The two most abundant isotopes of potassium are the stable
$^{39}\mathrm{K}$ and $^{41}\mathrm{K}$ isotopes, with 93.7\% and
6.7\% abundance respectively. The next most abundant isotope is
$^{40}\mathrm{K}$ with a very long half life of $1.248\times10^9$
years and abundance of
 0.0117\%. This isotope is the most important source of
 radioactivity for humans.

\begin{figure}[b]
\begin{center}
  \includegraphics[width=10cm,clip]{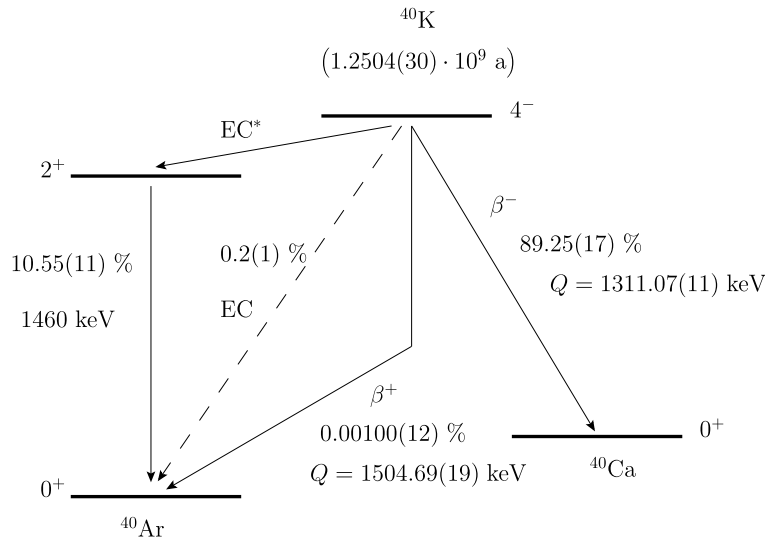}
\caption{Sketch showing the different decays and branching of
$^{40}$K. Reproduced with permission from:
  Pradler, J., Singh, B., Yavin, I.: On an unverified nuclear
decay and its role   in the {DAMA} experiment.  Phys. Lett. B
\textbf{720}(4-5), 399-.404 (2013). Copyright Creative Commons BY
3.0}
    \label{crowdion-figure15}    
\end{center}
 \end{figure}

As shown in Fig.~\ref{crowdion-figure15} and
table~\ref{crowdion-table:K40decay}, the nucleus $^{40}\mathrm{K}$
experiences decay through different branches with two daughter
nuclei $^{40}\mathrm{Ca}$ and
$^{40}\mathrm{Ar}$~\cite{crowdion-cameron2004,crowdion-radionuclides2012}.
The main parameters of the decay are $I$, the intensity of a given
branch in \% and $Q$, which is the difference between the rest
masses of the parent and daughter {\em atoms}. The difference of
mass between atoms is better tabulated than between nuclei. As the
atoms are neutral the mass difference between nuclei has to take
into account the difference in the number of electrons in the
neutral atoms. The available energy will depend on the rest mass
of the parent and daughter nuclei and other particles. It will be
obtained below for each type of decay.

The decay branches, $\beta-$ and $\beta+$  involve the emission of
an electron or a positron and a neutrino. The electron or positron
velocities are such that they have to be treated relativistically,
while the recoil velocity of the much heavier nuclei can be
described classically. We will suppose an electron to simplify the
language, but a positron can be equally  described  in what
follows. The maximum recoil energy of the nucleus is obtained when
the neutrino gets no kinetic energy. The recoil energy is much
smaller than the electron energy, so it can be neglected in the
energy calculations while due to its large mass, it is essential
for the momenta balance. The electron maximum energy is
$E_e=m_e\,c^2+E$, being $E$ the available energy in the decay, and
$E_e^2=m_e^2\,c^4+p_e^2c^2$, where $p_e$ is the momentum of the
electron. Considering the parent nucleus at rest, the momentum of
the nucleus is identical to the momentum of the electron
$p_N=p_e=(1/c)(E_e^2-m_e^2c^4)^{1/2}$ and the maximum nucleus
recoil energy is given by $E_N=p_N^2/(2 m_N)$. The decays always
involve the emission of a neutrino and may include the emission of
a photon, either $\gamma$ from the nucleus of X from the electron
shell, although the latter have much smaller energy and momentum
and will be of no importance for K$^+$\, recoil. The neutrino can
be considered as a massless particle as its rest mass it known to
be below $2.2$\,eV/c$^2$. Therefore for photons or neutrinos their
energy is given by $E_{\nu,\gamma,X}=p c$. If only a photon or a
neutrino is emitted the recoil momentum $p_N$ is equal to the
momentum of the photon or neutrino and trivially
$E_N=P_N^2/(2\,m_N)$. If there are only two daughter particles the
recoil energy $E_N$ has a single value.

Other data of interest are the ionization energies of K and of the
daughter nuclei.  If the recoil energy is larger than the
ionization energy of the atoms that interact, it can be used to
ionize an atom or ion and the energy cannot be transferred to the
neighbours. The ionization energies of the daughter atoms from
$^{40}$K decay can be seen in
table~\ref{crowdion-table:ionization}. A examination of the
possible ionization processes is done in the following subsection.

\begin{table}
\begin{center}
\caption{Table of ionization energies (eV) of atoms involved in
$^{40}$K decay~\cite{crowdion-lide2010}}
\label{crowdion-table:ionization}
\begin{tabular}{p{2cm} p{1.8cm} p{1.8cm} p{1.8cm} p{1.8cm} p{1.8cm}}
\hline\noalign{\smallskip}
Element & I & II & III & IV & V  \\
\noalign{\smallskip}\svhline\noalign{\smallskip}
Ar & 15.76 & 27.63  & 40.74 & 59.81 & 75.02\\
K  & ~~4.34 & 31.63  & 45.81 & 60.91 & 82.66\\
Ca & ~~6.11  & 11.87  & 50.91 & 67.27 & 84.50 \\
\hline\noalign{\smallskip}
\end{tabular}
\end{center}
\end{table}
           \rcgindex{\myidxeffect{I}!Ionization energy (Ar) }
           \rcgindex{\myidxeffect{I}!Ionization energy (Ca) }
                      \rcgindex{\myidxeffect{I}!Ionization energy (K) }
As the lattice is formed by  K$^+$\,, it is probable that the
second ionization of K, 31.6\,eV, is an upper limit for crowdions
or single row kink energies.

\subsection{$^{40}\mathrm{K}$ decay branches}
           \rcgindex{\myidxeffect{K}!$^{40}$K decay branches}

Here, we analyze in detail the different decay branches. A summary
if presented in table~\ref{crowdion-table:K40decay}.

The $^{40}\mathrm{K}$ decay branch that leads to
$^{40}\mathrm{Ca}$ is:

\begin{description}[]
\item[{\em $\beta^-$ decay}:]\mbox{}\\
\item[$\beta^-$: {\em  Decay with emission of an electron}.]\mbox{}\\
With $I\beta^-=89.25\%$ and mass difference between atoms
$Q=1311.07$\,keV~\cite{crowdion-radionuclides2012}. As the Ca atom
has an extra electron, discarding the electron binding energy of a
few keV, the mass difference between nuclei is $Q+m_e c^2$ and the
energy available when emitting an electron is $E\simeq Q+m_e
c^2-m_e c^2\simeq Q$ which will be shared between the electron and
the antineutrino emitted. Therefore, the maximum kinetic energy of
the electron or {\em endpoint} is almost equal to $Q$. The
daughter nuclei of $^{40}\mathrm{Ca}$ have a continuous
distribution of energy with a maximum of $E_k=42$\,eV at the
endpoint corresponding to a velocity $V=14.4$\,km/s.
 \rcgindex{\myidxeffect{K}!$^{40}$K $\beta^-$ decay}
 \rcgindex{\myidxeffect{B}!$\beta^-$ decay ($^{40}$K)}

The proton number increases by one, but the number of electrons
does not change, therefore the daughter ion would be
$\mathrm{Ca}^{++}$ with $50.6$\,eV third ionization energy.  This
is a likely origin of quodons for the decays with recoil energy
smaller than the 31.6\,eV K second ionization energy. The recoils
with larger energy will be able to deliver up to 10.4\,eV  after
the first collision that could produce breathers but not
crowdions.

\end{description}
The following processes have $^{40}\mathrm{Ar}$ as daughter nuclei
being the difference between the {\em atomic} masses
$Q=1504.69$\,keV. As the Ar atom has an electron less than K,
discarding the electron binding energies the mass difference
between nuclei is $\simeq Q-m_e c^2$ and the energy available
depends on the specific decay.

\begin{description}[]
\item[{EC1: \em Electron capture with decay to $^{40}\mathrm{Ar}$ excited state and $\gamma$
radiation.}]
\mbox{}\\
 \rcgindex{\myidxeffect{K}!$^{40}$K electron capture}
 \rcgindex{\myidxeffect{E}!Electron capture ($^{40}$K)}
With $I\epsilon=10.55\%$,  an electron from the shell is captured,
therefore the available energy is $E\simeq Q-m_ec^2+m_ec^2\simeq
Q$. In this decay a monoenergetic neutrino of 44\,keV is emitted
with negligible recoil (26\,meV) and the daughter nucleus is in an
excited state. Thereafter, the excited nucleus decays to the
ground state with the emission of a 1460\,keV $\gamma$
ray~\cite{crowdion-radionuclides2012}. The corresponding K$^+$\,
recoil energy of $^{40}\mathrm{Ar}$ is $E_k\simeq 29.2$\,eV with
velocity $V=12.0$\,km/s. As this is a two body process $E_k$ has
only slight variations due to interactions with the shell
electrons.
          \rcgindex{\myidxeffect{G}!$\gamma$ emission ($^{40}$Ar)}
           \rcgindex{\myidxeffect{A}!$^{40}$Ar $\gamma$ emission}

As no charge is emitted from the ion K$^+$\,, the daughter will
also be a monovalent ion of $\mathrm{Ar}^{+}$, with 27.7\,eV
second ionization energy. So there is  some probability that the
first $\mathrm{Ar}^{+}$ collision with K$^+$\, will further ionize
$\mathrm{Ar}^{+}$. The remaining energy 1.3 eV will not be enough
to produce a kink but may produce a breather.

\item[EC1+CE: {\em Electron capture with decay to $^{40}\mathrm{Ar}$
excited state and conversion electron.}] This is actually a subset
of the previous decay, but with a probability I=0.001 per 100
decays, the 1460\,keV $\gamma$ ray emitted can interact with the
shell and deliver the energy to an electron, called a conversion
electron. Save for a few keV of binding energy the $\gamma$ energy
is converted into kinetic energy of the electron, with a recoil
for the ion of 49.7\,eV and 15.6\,km/s. This is the largest energy
of all the recoils.
 \rcgindex{\myidxeffect{K}!$^{40}$K electron conversion}

 \rcgindex{\myidxeffect{E}!Electron conversion ($^{40}$K)}

As an electron has been emitted from the shell, the daughter ion
will be $\mathrm{Ar}^{++}$ with 40.8\,eV third  ionization energy.
This ionization and the 31.6\,eV second one of K$^+$\,  are likely
to occur. The remaining  energies of 8.8 or 18\,eV cannot produce
a crowdion but will be able to produce breathers.

\item[EC2: {\em Electron capture with direct decay to
$^{40}\mathrm{Ar}$ ground state}.]\mbox{}\\
With probability $I=0.2\%$, the energy available as in the decay
above is $E\simeq Q= 1504.69$.  There is a direct decay to the
ground state of $^{40}$Ar after the capture of a shell electron
and the emission of a monoenergetic neutrino that takes most of
the energy available $E \simeq 1504.69$\,keV minus the electron
binding energy which is only a few
keV\,\cite{crowdion-pradler2013,crowdion-sinev2015}. The recoil
energy is 31.1\,eV. The shell emits a 3\,keV Auger electron when
another electron of the shell occupies the vacancy left by the
captured electron, however, this has a negligible recoil.
 \rcgindex{\myidxeffect{K}!$^{40}$K direct decay to $^{40}$Ar}
 \rcgindex{\myidxeffect{A}!$^{40}$Ar ground state (decay to)}

The daughter nucleus has lost a positive unit charge but also the
shell has lost two electrons, the captured one plus the Auger
electron. Therefore the daughter ion will be Ar$^{++}$, which has
too little  energy for further ionization of Ar$^{++}$ or K$^{+}$
which need 40.8 and 31.6\,eV, respectively. Therefore, it would be
a likely source of crowdions but difficult to distinguish from the
$\beta^-$ recoil.

\item[$\beta^+$: {\em  decay with positron emission}.] \mbox{}\\
With very low probability $I\beta^+=0.001\%$, the available energy
is the mass difference between nuclei minus the mass energy of the
positron emitted, that is, $E\simeq Q-m_ec^2-m_ec^2=Q-2 m_e
c^2=483.7\,$keV. The energy $E$ is shared between a neutrino, the
emitted positron and the daughter nucleus. Therefore, the
positrons have a continuum of  energies with a maximum one or
endpoint
483.7\,keV~\cite{crowdion-engelkemeir1962,%
crowdion-cameron2004,crowdion-radionuclides2012}, which leads to
the maximum recoil energy $E_k\simeq 10$\,eV and velocity of
7\,km/s.
 \rcgindex{\myidxeffect{K}!$^{40}$K $\beta^+$ decay}
 \rcgindex{\myidxeffect{B}!$\beta^+$ decay ($^{40}$K)}
 \rcgindex{\myidxeffect{K}!$^{40}$K positron emission}
 \rcgindex{\myidxeffect{E}!Emission of positrons ($^{40}$K)}
\begin{table}[b]
\begin{center}
\caption{Table of decays for $^{40}$K}
\label{crowdion-table:K40decay}
\begin{tabular}{p{3.5cm} p{1.5cm} p{1.5cm} p{1.5cm} p{1.5cm} p{1.5cm}}
\hline\noalign{\smallskip}
Decay & $\beta^-$ & EC1 & EC1+CE$^1$ & EC2$^2$ & $\beta^+$\\
\noalign{\smallskip}\svhline\noalign{\smallskip}
Intensity & 89.25\% & 10.55\% & 0.001\% & 0.2\% & 0.001\%\\
T (keV)& 1311.07& 1460& 1460& 1504.69&483.7\\
Emitted charged particle& e$^{-}$& \mbox{} & e$^{-}$& e$^{-}$ & e$^{+}$\\
Recoil from & $\nu$+e$^{-}$& $\gamma$ & $e^{-}$ & $\nu$ & $\nu$+e$^{+}$\\
Max Recoil (eV)& 42& 29.2$^M$& 49.7$^M$& 31.1$^M$& 10\\
Daugther ion (A=40) & Ca$^{++}$& Ar$^{+}$& Ar$^{++}$& Ar$^{++}$& Ar\\
Max V (Km/s)& 14.4& 12$^M$& 15.7$^M$& 12.2$^M$&7\\
Ionization of daughter (eV)& 50.6& 27.7& 40.8& 40.8&15.8\\
$\Delta$q (e)& +1& 0& +1& +1&-1\\
\hline\noalign{\smallskip}
\end{tabular}
\begin{tabular}{p{11cm}}
 $^1$ Subset of EC1 when the gamma is delivered to a shell
 electron; \hspace{1cm}
 $^M$ Monocromatic\\
$^2$ Direct decay to Ar ground state, recoil from neutrino
emission; 3\,KeV Auger e$^{-}$\\
EC: electron capture; CE: conversion electron; T: energy available
excluding rest masses \\
Ionization energy of K$^{+}$  31.6\,eV\\
\hline\noalign{\smallskip}
\end{tabular}
\end{center}
\end{table}

As the atomic number is decreased by one unit to $Z-1$, the
initial ion K$^+$\, has lost a positive unit charge, but there has
been no change in the number of electrons, thus the daughter ion
will be a neutral Ar interacting with short range forces with the
neighbouring K$^+$. The first ionization energy of Ar is 15.8\,eV,
so, actually, the Ar atom has less of the required energy for
ionizing itself or for further ionization of K$^+$\, and will be
able to keep the 10\,eV energy. This seems too little to produce a
kink but may produce a breather. Due to their positive charge,
positrons leave tracks  in mica
muscovite~\cite{crowdion-russell88b,crowdion-steeds93}.
\end{description}

A study with the correlation of positron tracks, thickness
distribution of quodon tracks and other characteristics could make
it  possible to confirm the nature and characteristics of quodons.
See {\em Tracks in mica: 50 years later} for more
details~\cite{crowdion-russelltracks2015}.

\subsection{Secondary processes}
 \rcgindex{\myidxeffect{E}!Electron--positron pair production}
\begin{description}[]
\item[{\em Electron--positron pair production}:]\mbox{}\\
This is a secondary process after the $\gamma$ ray emission of
1460.82\,keV considered above~\cite{crowdion-engelkemeir1962}. It
needs the interaction of the $\gamma$ ray with a nucleus, and the
produced positron and electron can share the energy in any
proportion. The maximum recoil energy corresponds to a single
particle taking almost all the energy except for the small amount
taken by the nucleus, which is necessary due to momentum
conservation. The available kinetic energy is
$E=E_\gamma-2\,m_e\,c^2=437.4$\,keV and the maximum recoil energy
is $E_k=8.8$\,eV.  The probability of the combined process of
electron capture and pair production is of the same order of
magnitude as $\beta^+$ emission and also the energies are
similar~\cite{crowdion-engelkemeir1962}. The probability of
interaction of the $\gamma$ ray with a nucleus is proportional to
$Z^2$ which favors the interaction with potassium; however,
potassium atoms are only 5\% of the atoms in mica.

As the energy is smaller than the second ionization energy of K of
31.6\,eV it is likely that the subsequent K$^+$\,--K$^+$\,
collisions are elastic.
\end{description}

Other secondary processes may also occur via other radioactive
nuclei and their corresponding decay, but it will be beyond the
    objective of this work to continue the subject further.

\section{Summary}
\label{sec:crowdion-summary} In this work we have described in
detail the magic mode for the strain or compression of the bonds.
A construction in terms of phasors has been developed in order to
obtain an intuition of the relative phases and behaviour of the
particles as the kink passes over them. We have considered an 1D
model for the close-packed lines of potassium ions inside a cation
layer of mica muscovite using realistic potentials. There exists
only a single kink with a specific velocity and energy dubbed the
crowdion. It is relatively well described by the magic mode but
the kink is transformed into a double kink. It leaves behind a
phonon wave with exponentially diminishing amplitude that travels
at the same velocity as that of the kink. Simulations with
different initial energies bring about a variety of phenomena
including the formation of two crowdions that leave behind
nonlinear waves and phonons. The crowdions also survive at
temperatures of 300-1000\,K. Finally, an analysis of the possible
decay modes of $^{40}$K has been performed including their
possible consequences with respect to crowdion formation.
 A careful study of the tracks in mica muscovite
compared with the decay modes can shed light on their
characteristics and origin.

The energy of the kinks or crowdions described in this work can be
provided by the $^{40}$K decay and is enough to expel an atom at
the border. The crowdions survive to high temperature and travel
long distances. They transport positive charge and therefore are
very likely to be recorded in the form of dark tracks in mica
muscovite. If they are the cause of the quodons or other marks
observed in this mineral is still an open question.

\section*{Acknowledgments}
J.F.R.A., V.S.M., and L.M.G.R. acknowledge financial support from
the projects FIS2008-04848, FIS2011-29731- C02-02, and
MTM2012-36740-C02-02 from Mi\-nisterio de Ciencia e
Innova\-ci\'on´ (MICINN). All authors acknowledge Prof. F.M.
Russell for ongoing discussions.

\bibliographystyle{spmpsci}  
\bibliography{crowdion3}

\begin{thebibliography}{10}
\providecommand{\url}[1]{{#1}}
\providecommand{\urlprefix}{URL }
\expandafter\ifx\csname urlstyle\endcsname\relax
  \providecommand{\doi}[1]{DOI~\discretionary{}{}{}#1}\else
  \providecommand{\doi}{DOI~\discretionary{}{}{}\begingroup
  \urlstyle{rm}\Url}\fi

\bibitem{crowdion-archilla2013}
Archilla, J.F.R., Kosevich, {\relax Yu}.A., Jim\'enez, N.,
  S\'{a}nchez-Morcillo, V.J., Garc\'{\i}a-Raffi, L.M.: Moving excitations in
  cation lattices.
\newblock Ukr. J. Phys. \textbf{58}(7), 646--656 (2013)

\bibitem{crowdion-archilla-springer2014}
Archilla, J.F.R., Kosevich, {\relax Yu}.A., Jim\'enez, N., S\'anchez-Morcillo,
  V.J., Garc\'{i}a-Raffi, L.M.: Supersonic kinks in {C}oulomb lattices.
\newblock In: R.~Carretero-Gonz\'alez, et~al. (eds.) Localized Excitations in
  Nonlinear Complex Systems, pp. 317--331. Springer, New York (2014)

\bibitem{crowdion-archilla-kosevich2015}
Archilla, J.F.R., Kosevich, {\relax Yu}.A., Jim\'enez, N., S\'anchez-Morcillo,
  V.J., Garc\'{\i}a-Raffi, L.M.: Ultradiscrete kinks with supersonic speed in a
  layered crystal with realistic potentials.
\newblock Phys. Rev. E \textbf{91}, 022,912 (2015)

\bibitem{crowdion-cameron2004}
Cameron, J.A., Singh, B.: Nuclear data sheets for {A}=40.
\newblock Nucl. Data Sheets \textbf{102}({2}), 293--513 (2004)

\bibitem{crowdion-CollinsAM92}
Collins, D.R., Catlow, C.R.A.: Computer simulation of structure and cohesive
  properties of micas.
\newblock Am. Mineral. \textbf{77}(11-12), 1172--1181 (1992)

\bibitem{crowdion-diaz2000}
Diaz, M., Farmer, V.C., Prost, R.: Characterization and assignment of far
  infrared absorption bands of {K}$^+$ in muscovite.
\newblock Clays Clay Miner. \textbf{48}, 433--438 (2000)

\bibitem{crowdion-durrani2008}
Durrani, S.A.: Nuclear tracks today: Strengths, weaknesses, challenges.
\newblock Rad. Meas \textbf{43}, S26--S33 (2008)

\bibitem{crowdion-engelkemeir1962}
Engelkemeir, D.W., Flynn, K.F., Glendenin, L.E.: Positron emission in the decay
  of {K}$^{40}$.
\newblock Phys. Rev. \textbf{126}(5), 1818--1822 (1962)

\bibitem{crowdion-fleischer2011}
Fleischer, R.: Tracks to Innovation. Nuclear Tracks in Science and Technology.
\newblock Springer, New York (2011)

\bibitem{crowdion-friesecke2002}
Friesecke, G., Matthies, K.: Atomic-scale localization of high-energy solitary
  waves on lattices.
\newblock Physica D \textbf{171}(4), 211 -- 220 (2002)

\bibitem{crowdion-gedeon2002}
Gedeon, O., Machacek, J., Liska, M.: Static energy hypersurface mapping of
  potassium cations in potassium silicate glasses.
\newblock Phys. Chem. Glass. \textbf{43}(5), 241--246 (2002)

\bibitem{crowdion-KdV1895}
Korteweg, D.J., de~Vries, F.: On the change of form of long waves advancing in
  a rectangular canal, and on a new type of long stationary waves.
\newblock Philosophical Magazine \textbf{39}, 422--443 (1895).
\newblock See also
  \url{http://en.wikipedia.org/wiki/Korteweg-de\_Vries\_equation}

\bibitem{crowdion-kosevich73}
Kosevich, A.M., Kovalev, A.S.: The supersonic motion of a crowdion. {T}he one
  dimensional model with nonlinear interaction between the nearest neighbors.
\newblock Solid State Commun. \textbf{12}, 763--764 (1973)

\bibitem{crowdion-kosevich93}
Kosevich, {\relax Yu}.A.: Nonlinear sinusoidal waves and their superposition in
  anharmonic lattices.
\newblock Phys. Rev. Lett. \textbf{71}, 2058--2061 (1993)

\bibitem{crowdion-kosevich04}
Kosevich, {\relax Yu}.A., Khomeriki, R., Ruffo, S.: Supersonic discrete
  kink-solitons and sinusoidal patterns with {\em magic} wave number in
  anharmonic lattices.
\newblock Europhys. Lett. \textbf{66}, 21--27 (2004)

\bibitem{crowdion-kudriavtsev2005}
Kudriavtsev, Y., Villegas, A., Godines, A., Asomoza, R.: Calculation of the
  surface binding energy for ion sputtered particles.
\newblock Appl. Surf. Sci. \textbf{239}(3-4), 273--278 (2005)

\bibitem{crowdion-lide2010}
Lide, D.R. (ed.): Handbook of Chemistry and Physics, 90$^{\mathrm{th}}$ edn.
\newblock {CRC} {P}ress, Boca Raton, Florida, USA (2010).
\newblock Section 10, page 203

\bibitem{crowdion-milchev90}
Milchev, A.: Breakup threshold of solitons in systems with nonconvex
  interactions.
\newblock Phys. Rev. B \textbf{42}, 6727--6729 (1990)

\bibitem{crowdion-moleron2014}
Moler\'on, M., Leonard, A., Daraio, C.: Solitary waves in a chain of repelling
  magnets.
\newblock J. Appl. Phys. \textbf{115}(18), 184,901 (2014)

\bibitem{crowdion-radionuclides2012}
Mougeot, X., Helmer, R.G.: {LNE-LNHB/CEA}--{T}able de {R}adionucl\'eides,
  {K}-40 tables.
\newblock {\tt http://www.nucleide.org } (2012)

\bibitem{crowdion-pradler2013}
Pradler, J., Singh, B., Yavin, I.: On an unverified nuclear decay and its role
  in the {DAMA} experiment.
\newblock Phys. Lett. B \textbf{720}(4-5), 399--404 (2013)

\bibitem{crowdion-pricewalker62}
Price, P.B., Walker, R.M.: Observation of fossil particle tracks in natural
  micas.
\newblock Nature \textbf{196}, 732--734 (1962)

\bibitem{crowdion-russell67a}
Russell, F.M.: The observation in mica of tracks of charged particles from
  neutrino interactions.
\newblock Phys. Lett. \textbf{25B}, 298--300 (1967)

\bibitem{crowdion-russell67b}
Russell, F.M.: Tracks in mica caused by electron showers.
\newblock Nature \textbf{216}, 907--909 (1967)

\bibitem{crowdion-russell88b}
Russell, F.M.: Identification and selection criteria for charged lepton tracks
  in mica.
\newblock Nucl. Tracks. Rad. Meas. \textbf{15}, 41--44 (1988)

\bibitem{crowdion-russellcrystal2015}
Russell, F.M.: I saw a crystal.
\newblock In: J.F.R. Archilla, N.~Jim\'enez, V.J. S\'anchez-Morcillo, L.M.
  Garc\'{i}a-Raffi (eds.) Quodons in mica: nonlinear localized travelling
  excitations in crystals. Springer (\mbox{}2015).
\newblock To appear

\bibitem{crowdion-russelltracks2015}
Russell, F.M.: Tracks in mica, 50 years later.
\newblock In: J.F.R. Archilla, N.~Jim\'enez, V.J. S\'anchez-Morcillo, L.M.
  Garc\'{i}a-Raffi (eds.) Quodons in mica: nonlinear localized travelling
  excitations in crystals. Springer (\mbox{}2015).
\newblock To appear

\bibitem{crowdion-russellcollins95b}
Russell, F.M., Collins, D.R.: Lattice-solitons and non-linear phenomena in
  track formation.
\newblock Rad. Meas \textbf{25}, 67--70 (1995)

\bibitem{crowdion-russellcollins95a}
Russell, F.M., Collins, D.R.: Lattice-solitons in radiation damage.
\newblock Nucl. Instrum. Meth. B \textbf{105}, 30--34 (1995)

\bibitem{crowdion-russelleilbeck07}
Russell, F.M., Eilbeck, J.C.: Evidence for moving breathers in a layered
  crystal insulator at 300\,{K}.
\newblock Europhys. Lett. \textbf{78}, 10,004 (2007)

\bibitem{crowdion-russelleilbeck2011}
Russell, F.M., Eilbeck, J.C.: Persistent mobile lattice excitations in a
  crystalline insulator.
\newblock Discret. Contin. Dyn. S.-S \textbf{4}, 1267--1285 (2011)

\bibitem{crowdion-savin95}
Savin, A.V.: Supersonic regimes of motion of a topological soliton.
\newblock Sov. Phys. JETP \textbf{81}(3), 608--613 (1995)

\bibitem{crowdion-schlosserrussell94}
Schl\"{o}{\ss}er, D., Kroneberger, K., Schosnig, M., Russell, F.M., Groeneveld,
  K.O.: Search for solitons in solids.
\newblock Rad. Meas \textbf{23}, 209--213 (1994)

\bibitem{crowdion-sinev2015}
Sinev, V.V., Bezrukov, L.B., Litvinovich, E.A., Machulin, I.N., Skorokhvatov,
  M.D., Sukhotin, S.V.: Looking for antineutrino flux from 40{K} with large
  liquid scintillator detector.
\newblock Phys. Part. Nuclei \textbf{46}(2), 186--189 (2015)

\bibitem{crowdion-steeds93}
Steeds, J.W., Russell, F.M., Vine, W.J.: Formation of epidote fossil positron
  tracks in mica.
\newblock Optik \textbf{92}, 149--154 (1993)

\bibitem{crowdion-ziegler2008}
Ziegler, J.F., Biersack, J.P., Ziegler, M.D.: SRIM - The Stopping and Range of
  Ions in Matter.
\newblock Published by James Ziegler, Chester, Maryland (2008)

\bibitem{crowdion-zolotaryuk97}
Zolotaryuk, Y., Eilbeck, J.C., Savin, A.V.: Bound states of lattice solitons
  and their bifurcations.
\newblock Physica D \textbf{108}, 81--91 (1997)

\end{thebibliography}
\end{document}